\documentclass[fleqn,usenatbib]{mnras}

\usepackage{newtxtext,newtxmath}

\usepackage[T1]{fontenc}
\usepackage{ae,aecompl}
\usepackage{siunitx}

\usepackage{graphicx}	
\usepackage{amsmath}	
\usepackage{amssymb}	
\usepackage{booktabs}
\usepackage{nccmath}
\usepackage{mathtools}
\usepackage{lastpage}
\usepackage{textgreek} 
\usepackage{makecell}
\usepackage{multirow}

\def\msun{{\rm M}_{\sun}}

\title[Calibrating cluster weak lensing]{Mapping dark matter and finding filaments:\\
calibration of lensing analysis techniques on simulated data
}

\author[Tam et al. 2020]{Sut Ieng Tam,$^{1}$\thanks{E-mail: sut-ieng.tam@durham.ac.uk}
Richard Massey,$^{1}$
Mathilde Jauzac$^{2,1,3}$
and
Andrew Robertson$^{1}$ 
\\
\\
\\
$^{1}$Institute for Computational Cosmology, Durham University, South Road, Durham DH1 3LE, UK\\
$^{2}$Centre for Extragalactic Astronomy, Durham University, South Road, Durham DH1 3LE, UK\\
$^{3}$Astrophysics and Cosmology Research Unit, School of Mathematical Sciences, University of KwaZulu-Natal, Durban 4041, South Africa\\
}

\date{Accepted ---. Received ---; in original form \today}

\begin{document}
\label{firstpage}
\pagerange{\pageref{firstpage}--\pageref{lastpage}}
\maketitle

\begin{abstract}
We quantify the performance of mass mapping techniques on mock 
imaging and gravitational lensing data of galaxy clusters.
The optimum method depends upon the scientific goal. We assess measurements
of clusters' radial density profiles, departures from sphericity, and their filamentary attachment to the cosmic web.
We find that mass maps produced by direct (KS93) inversion of shear measurements are unbiased, and that their noise can be suppressed via filtering with {\sc MRLens}. Forward-fitting techniques, such as {\sc Lenstool}, suppress noise further, but at a cost of biased ellipticity in the cluster core and over-estimation of mass at large radii.
Interestingly, current searches for filaments are noise-limited by the intrinsic shapes of weakly lensed galaxies, rather than by the projection of line-of-sight structures. Therefore, space-based or balloon-based imaging surveys that resolve a high density of lensed galaxies, could soon detect one or two filaments around most clusters.
\end{abstract}




\section{Introduction}

The {\textLambda}CDM standard model of cosmology suggests that structures in the Universe formed hierarchically, via mergers of small over-densities in the early Universe into larger and larger objects \citep{1978MNRAS.183..341W,Springel2005SimulationsOT,Schaye:2014tpa}. Thirteen billion years after the Big Bang, the largest objects are currently clusters of hundreds or thousands of galaxies. Because their growth has spanned the entire age of the Universe, and has depended upon the density of building material and its collapse under gravity, versus its disruption by supernovae, active galactic nuclei, and dark energy, measurements of the precise number and properties of clusters is a highly sensitive test of the standard cosmological model\ \citep[e.g.][]{1993ApJ...407L..49B,2003ApJ...588L...1B,Ho:2005ea,2010ApJ...708..645R,2015PNAS..11212249W,Jauzac:2016tjc,Schwinn2017,2018MNRAS.478L..34M,Fluri:2019qtp}.

Gravitational lensing is particularly efficient at investigating clusters. The dense concentration of mass in a foreground galaxy cluster deflects light rays emitted by unrelated galaxies far in the background. Since adjacent light rays are almost coherently deflected, the shapes of those distant galaxies appear distorted, and typically stretched in such a way that their long axes make circular patterns around the cluster. Crucially, the deflection of light rays depends only upon the total projected mass distribution. Measurements of gravitational lensing are therefore uniquely sensitive to the distribution of invisible-but-dominant dark matter, and unbiased by the nature and dynamical state of ordinary matter\ \citep[e.g.][]{hrev,mrev,knrev,trev,kirev,srev}.

Ground-based observations of gravitational lensing by galaxy clusters have been successfully used to measure clusters' average or bulk properties, such as mass \citep[e.g.][]{2014MNRAS.439....2V,Umetsu:2014vna,2016MNRAS.461.3794O,2018PASJ...70S..28M,2017MNRAS.472.1946S,Schrabback:2016hac,McClintock:2018bxh,Miyatake:2018lpb,Rehmann:2018nis,2020ApJ...890..148U,2019arXiv191204414H}, and ellipticity \citep[e.g.][]{Evans:2008mp,2010MNRAS.405.2215O,Clampitt:2015wea,vanUitert:2016guv,Shin:2017rch,Umetsu:2018ypz,Chiu:2018gok}.
The CLASH survey \citep[Cluster Lensing and Supernova Survey with Hubble;][]{2012ApJS..199...25P} measured the mass and concentration of 25 clusters, by combining wide-field Subaru imaging with \emph{Hubble Space Telescope} (\emph{HST}) imaging of the cluster cores \citep{2015ApJ...806....4M}.
However, ground-based observations have yielded only marginally significant detections of filaments  \citep[e.g.][]{2006A&A...451..395C,Kaiser:1998ja,2002ApJ...568..141G,2004A&A...422..407G,2012Natur.487..202D,Martinet:2016ind}, whose dark matter density is too low (and the filaments too narrow to resolve).

Space-based imaging reveals the shapes of more background galaxies, and increases the S/N of lensing measurements in multiple resolution elements across an individual cluster. Thus the shape and morphology of individual mass distributions can be precisely {\em mapped}, without the need to average out features over a population of clusters. Space-based lensing reconstructions have resolved substructure near cluster cores \citep[e.g.][]{2011MNRAS.417..333M,Natarajan:2017sbo}; bimodality even in relatively distant clusters like the `Bullet Cluster' \citep{Bradac:2006er} or `El Gordo' \citep{2014ApJ...785...20J}; and filaments in Abell\,901/902 \citep{2008MNRAS.385.1431H} and MACSJ\,0717+3745 \citep{2012MNRAS.426.3369J}. Nonetheless, these analyses remain rare because the $\sim3\arcmin\times3\arcmin$ field of view of \emph{HST}'s Advanced Camera for Surveys (ACS) is smaller than a typical cluster's angular size. Furthermore, both of \emph{HST}'s contiguous surveys (GOODS and COSMOS) unluckily sampled regions of the Universe that are underdense at the $z=0.2$--$0.4$ redshifts where lensing is most sensitive \citep[][]{Heymans:2004zp,Massey:2007gh,Krolewski:2017jsm}, so happen to contain few lensing clusters \citep{2007ApJS..172..254G,2007Natur.445..286M}. Until recently, only around one cluster, MS\,0451-03, had a dedicated wide-field mosaic of contiguous \emph{HST} imaging had been obtained \citep{2007ApJ...671.1503M}.

There will soon be wide-field, space-resolution imaging taken around 6 more clusters through the \emph{HST}/BUFFALO survey \citep{2020ApJS..247...64S}, 200 more clusters from the balloon-borne telescope \emph{SuperBIT} \citep{2016arXiv160802502R,2018SPIE10702E..0RR}, and 10,000 from \emph{Euclid} \citep{2011arXiv1110.3193L}. In the next decade, 40,000 clusters will be observed to even greater depth by \emph{WFIRST} \citep[Wide Field Infrared Survey Telescope;][]{2013arXiv1305.5422S}. 

The intent of this work is to prepare for future observations, much as \cite{2013MNRAS.433.3373V} calibrated mass mapping methods for the current generation of wide-field ground-based lensing surveys.
We use mock space-based weak-lensing data to develop and quantify the performance of two different methods to map dark matter around galaxy clusters, to measure deviations from sphericity, and to search for filaments connecting it with the cosmic web. 
Where we must make decisions about general properties (e.g.\ distance, mass) of clusters that we simulate, we shall use MS\,0451-03 as a template, so our predictions can be immediately tested on real observations \citep[see our companion paper,][]{2020arXiv200610156T}.
 
This paper is organised as follows. We summarise background theory in Section~\ref{sec:theory}, and introduce the simulated data in Section~\ref{sec:data}. In the context of various scientific motivations, we describe weak-lensing mass mapping and analysis techniques in Section~\ref{sec:method}. We quantify their results in Section~\ref{sec:results}, and conclude in Section~\ref{sec:conclusion}. Throughout the paper, we define angular diameter distances assuming a background cosmology with $\Omega_m=0.287$, $\Omega_{\Lambda}=0.713$, and $h=H_0/100 \si{kms^{-1}Mpc^{-1}}=0.693$  \citep[WMAP 9-year cosmology;][]{2013ApJS..208...19H}. All magnitudes are quoted in the AB system.

\section{Weak Gravitational Lensing Theory}
\label{sec:theory}

\subsection{Coherent deflection of light rays}
\label{sec:theory:deflection}
Gravitational lensing is the deflection of light rays from a distant source, by massive objects along our line of sight. 
The apparent shape of the source becomes distorted when a bundle of light rays from it are coherently distorted. 
Because cosmological distances are so large, the 3D distribution of intervening mass can be conveniently represented (through the `thin lens' approximation) as a 2D surface density, $\Sigma(\boldsymbol{R})$, where ${\boldsymbol{R}}=(x,y)$ is the 2D angular position in the plane of the sky. 
A similar projection can be applied to obtain a 2D effective gravitational potential $\varphi(\boldsymbol{R})$. 
The angle through which light rays are deflected corresponds to spatial derivatives in the gravitational potential.

In the weak-lensing regime, where deflection angles are small, the image distortions can be split into two dominant components. 
The first is an isotropic magnification, by a factor proportional to the projected density and known as `convergence'
\begin{equation}
\kappa(\boldsymbol{R})=\frac{\Sigma(\boldsymbol{R})}{\Sigma_c},
\label{eq:kappa}
\end{equation}
where the `critical density'
\begin{equation}
\Sigma_c=\frac{c^2}{4\pi G}\frac{D_s}{D_lD_{ls}}=\frac{c^2}{4\pi G D_l}\beta^{-1}(z_l,z_s), 
\end{equation}
depends upon the angular diameter distances from the observer to the lens, $D_l$, from the observer to the source, $D_s$, and from the lens to the source, $D_{ls}$. The lensing sensitivity function, $\beta(z_l,z_s)=D_{ls}/D_s$, describes the lensing  strength as a function of the lens and source redshifts ($z_l, z_s$). For a foreground galaxy with $z_s<z_l$, $\beta(z_l,z_s)=0$.
The second component of the distortion is a shear
\begin{equation}
\boldsymbol{\gamma}=\gamma_1+i\gamma_2=\lvert\boldsymbol{\gamma}\rvert e^{2i\phi},
\label{eq:gamma_ini}
\end{equation}
where the real component, $\gamma_1$, represents elongation along the $x$ direction, and the complex component, $\gamma_2$, represents elongation at $45^\circ$.

An observable quantity, `reduced shear'
\begin{equation}
\boldsymbol{g}\equiv\frac{\boldsymbol{\gamma}}{1-\kappa}
\label{eq:reducedshear}
\end{equation}
can be measured from the apparent shapes of galaxies.
In the weak-lensing regime, it is typically true that $\kappa\ll 1$, hence $\boldsymbol{g}\approx\boldsymbol\gamma$. For more information, see e.g.\  \cite{srev}.

\subsection{Analytic mass distributions}
\label{sec:theory:analytic}

In several places throughout this paper, we will approximate a mass distribution using one of two parametric models.
The models are usually described in circularly symmetric form, $\Sigma(|\boldsymbol{R}|)$ or $\varphi(|\boldsymbol{R}|)$, but can be made elliptical by a coordinate transformation 
\begin{equation}
  |\boldsymbol{R}'|^2=q(x^2\cos^2\phi+y^2\sin^2\phi)+(y^2\cos^2\phi-x^2\sin^2\phi)/q~,
\end{equation}
\citep{1993ApJ...417..450K,2010MNRAS.405.2215O} that maps a circle to an ellipse 
with axis ratio $0<q\leq 1$ and orientation $\phi$.
Except where mentioned explicitly, we apply this transformation to the projected mass distribution. 
Applying it instead to the gravitational potential yields different results, and no simple mapping exists between them. 

\subsubsection{tPIEMD profile}
\label{sec:theory:piemd}

Massive elliptical galaxies are empirically observed to have an approximately isothermal density distribution ($\rho\propto r^{-2})$, and total mass proportional to the velocity distribution of their stars, $\sigma$.
This would have an inconvenient mathematical singularity at the centre, which is removed in the truncated Pseudo-Isothermal Elliptical Mass Distribution (tPIEMD; \citealt{1993ApJ...417..450K,2005MNRAS.356..309L,eliasdottir2007}) 
\begin{equation}
\rho_\mathrm{tPIEMD}=\frac{\rho_0}{(1+r^2/r_\mathrm{c}^2)(1+r^2/r_\mathrm{t}^2)}.
\end{equation}
This has constant density 
\begin{equation}
     \rho_0=\frac{\sigma^2}{2\pi G}\frac{r_\mathrm{c}+r_\mathrm{t}}{r_\mathrm{c}^2~ r_\mathrm{t}}\,.
\end{equation}
inside core radius $r_\mathrm{c}$ and has finite integrated mass because of the truncation at radius $r_\mathrm{t}$. 
The projected two-dimensional mass distribution is
\begin{equation}
\Sigma_\mathrm{tPIEMD}({R})= \frac{\sigma^{2}}{2G} \frac{r_\mathrm{t}}{r_\mathrm{t}-r_\mathrm{c}} \left( \frac{1}{\sqrt[]{R^{2} + r_\mathrm{c}^{2}}} - \frac{1}{\sqrt[]{R^{2} + r_\mathrm{t}^{2}}} \right) ~.
\label{eq:PIMED}
\end{equation}

\subsubsection{NFW profile}
\label{sec:theory:nfw}

Numerical simulations suggest that the distribution of dark matter in isolated haloes forms a Navarro-Frenk-White \citep[NFW;][]{Navarro:1995iw,Navarro:1996gj} profile 
\begin{equation}
\rho_{\rm{NFW}}=\frac{\rho_\mathrm{s}}{(r/r_\mathrm{s})(1+(r/r_\mathrm{s}))^2}
\end{equation}
where $\rho_\mathrm{s}$ and $r_\mathrm{s}$ are a characteristic density and radius. 
For any given cosmology and cluster redshift, this model can also be parameterized in terms of a concentration $c_{200}\equiv r_{200}/r_s$, where $r_{200}$ is the 3D radius within which the mean enclosed density is equal to 200 times the critical density $\rho_{c}$ of the Universe, and halo mass $M_{200}\equiv (4\pi/3) 200\rho_{c}r_{200}^3$.
The projected two-dimensional mass distribution \citep{1996A&A...313..697B} is 
\begin{equation}
    \Sigma_\mathrm{NFW}(R)=2\rho_{\rm{s}}r_{
\rm{s}}F(x),
\label{eq:NFW}
\end{equation}
where  $x=R/r_{\rm{s}}$ and 
\begin{equation}
 F(x) = 
  \begin{cases} 
   \frac{1}{x^2-1}\left(1-\frac{2}{\sqrt{x^2-1}}\text{arctan}\sqrt{\frac{x-1}{x+1}}\right) & \text{if } x > 1 \\
   \frac{1}{3}       & \text{if } x = 1\\
   \frac{1}{x^2-1}\left(1-\frac{2}{\sqrt{1-x^2}}\text{arctan}\sqrt{\frac{1-x}{1+x}}\right) & \text{if } x < 1~.
  \end{cases}
\end{equation}

\section{Data}
\label{sec:data}

\begin{figure*}
\centering
\includegraphics[width=0.95\textwidth]{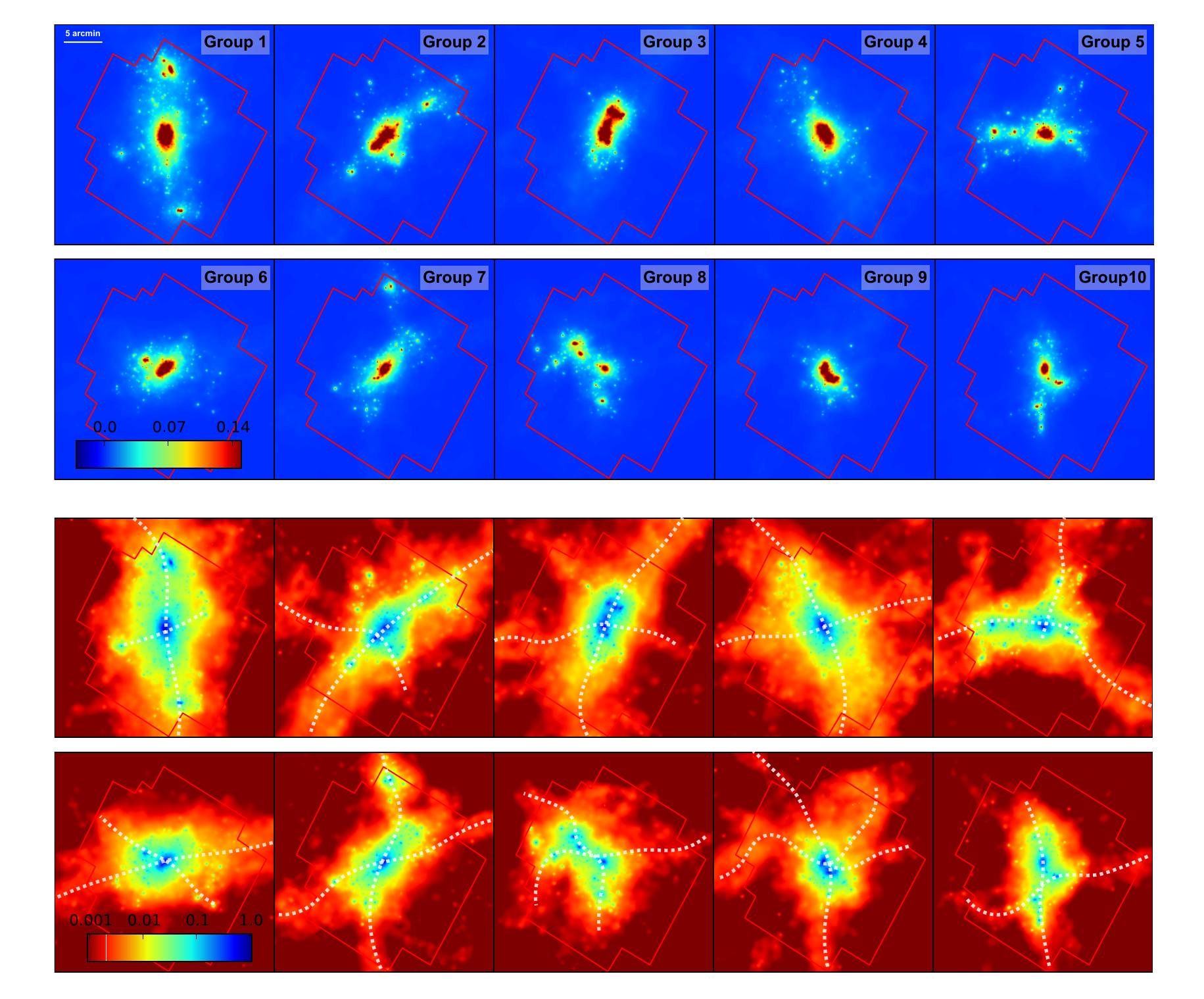}
\caption{Noise-free maps of the total mass distribution in the ten most massive clusters of the BAHAMAS simulations, projected along a randomly-oriented line of sight. 
Clusters have masses $M_{200}$ from $\num{2e15}\msun$ (cluster 1) to $\num{4e14}\msun$ (cluster 10), and are sorted in descending order of $M_{\rm FOF}$, as in Table~\ref{tab:clustermass}.
Colours show the lensing convergence $\kappa$ (Top panel: linear scale; Bottom panel: logarithmic scale). Dotted white lines show filaments identified from the noise-free, projected mass distribution, above density thresholds defined in section~\ref{sec:data:filaments}. 
For reference, red lines indicate the field of view in which \emph{HST} observations exist for real cluster MS\,0451-03. }
\label{fig:truekappa}
\end{figure*}

We use $N$-body particle data from the BAHAMAS suite of cosmological simulations \citep{2017MNRAS.465.2936M, 2018MNRAS.476.2999M}. These were run with different background cosmologies and implementations of sub-grid galaxy formation physics, and designed to test the impact of baryonic physics on large-scale structure tests of cosmology. For this paper, we use the version with a WMAP 9-year \citep{2013ApJS..208...19H} cosmology, and sub-grid feedback model that is calibrated to produce a good match to the observed stellar mass function, X-ray luminosities and gas fractions of galaxy clusters. This simulation occupies a periodic cubic volume, $400 \, h^{-1} \, \mathrm{Mpc}$ on a side, with dark matter and (initial) baryon particle masses of $5.5 \times 10^{9} \, \msun$ and $1.1 \times 10^{9} \, \msun$, respectively. 

\subsection{Distribution of mass in clusters}
\label{sec:data:clusters}

We extract the ten most massive clusters from the $z=0.5$ simulation snapshot. We first use the friends-of-friends algorithm (FOF; \citealt{2011ApJS..195....4M}) to identify all matter overdensities. For each FOF group, we calculate $r_{200}$ and $M_{200}$, the total mass enclosed within this sphere. For the ten most massive clusters, which have $\num{4e14}\msun < M_{200} < \num{2e15}\msun$, we store the 3D distribution of dark matter, stars and gas.

To generate a 2D, pixellated convergence map, we follow the method of \cite{2019MNRAS.488.3646R}. In summary, we project the location of all simulation particles within $5 \, r_{200}$ of the centre of a cluster along a line of sight (here, the simulation $z$-axis). In a $25\times25$\,Mpc ($2048\times2048$ pixel) map centred on the most bound particle, we use an adaptive triangular shaped cloud scheme to smooth each particle's mass over a kernel whose size depends on the 3D distance to that particle's 32nd nearest neighbour. 
Resulting convergence maps are shown in figure~\ref{fig:truekappa}, adopting the lens redshift $z_l = 0.55$ of galaxy cluster MS0451-03 as a concrete example, and source redshift $z_s = 0.97$ typical of \emph{HST} observations to single-orbit depth \citep{2007ApJS..172..219L}. The masses of the clusters are listed in Table~\ref{tab:clustermass}.

\begin{table}
\centering
\begin{tabular}{ccc}
\hline
\hline
 & $M_{{\rm{FOF}}}(10^{14}\msun)$ & $M_{{200}}(10^{14}\msun)$ \\
\hline
Cluster 1 & 27.7 & 17.3 \\
Cluster 2 & 17.9 & 15.0  \\
Cluster 3 & 17.8 & 17.7\\
Cluster 4 & 16.6 & 14.6 \\
Cluster 5 & 14.3 & 9.7 \\
Cluster 6 & 13.3 & 11.0 \\
Cluster 7 & 12.9 & 8.9 \\ 
Cluster 8 & 11.1 & 4.0 \\ 
Cluster 9 & 9.4 & 8.2 \\ 
Cluster 10 & 9.3 & 5.7 \\ 
\hline
\hline
\end{tabular}
\caption{Masses of the 10 most massive clusters in the BAHAMAS simulations, which we use as mock data for this study. Columns list the friends-of-friends masses $M_{\rm FOF}$, and overdensity mass $M_{200}$.}
\label{tab:clustermass}
\end{table}

\label{sec:data:filaments}

Before proceeding further, we identify 40 filaments in the ten projected mass maps, defined as radially extended regions with convergence $0.005<\kappa<0.01$, which is equivalent to a surface density of $1.7\times10^{7}<\Sigma\,(\rm{\msun/kpc^2})<3.4\times10^{7}$. These are indicated by white dashed lines in the bottom panel of figure~\ref{fig:truekappa}.

\subsection{Distribution of all other mass along a line of sight}
\label{sec:data:lss}

In addition to the mass of the galaxy cluster itself, we also account for large-scale structure (LSS) projected by chance along the same line of sight. 
This is a source of noise in the projected mass of the cluster, which is then added to the mock data in section~\ref{sec:data:mockshear}.

To quantify the expected level of noise, we generate realisations of LSS along 1000 random lines of sight through the BAHAMAS simulation box.
We then integrate the 3D mass along the line of sight, weighted by the lensing sensitivity function $\beta(z)$ with $\langle z_s\rangle=0.97$, interpreting it as a mass distribution in a single lens plane at $z_l=0.55$.
For each realisation of LSS, we calculate an effective radial density profile, $\kappa(R)$. 
The mean of these realisations is (unsurprisingly) consistent with zero; we also calculate the rms scatter $\sigma_{\rm{LSS}}$.
In concentric annuli of width $\Delta R=25\arcsec$, these are well-fit by 
\begin{equation}
    \label{eq:sigmaLSS}
    \sigma_{\rm{LSS}}(R)=\frac{A}{\sqrt{R(\rm{arcsec})}+B},
\end{equation}
with best-fit values for free parameters 
\begin{equation}
    A=0.197\pm0.008, \quad  B=6.441\pm0.502\, .
\end{equation}
We add this in quadrature to the statistical uncertainty on the reconstructed density profiles in Sect.~\ref{sec:method:radial}.
Note that it would also be possible to compute the full covariance matrix between LSS at different radii or in adjacent pixels of a mass map. 
Here we use only the diagonal elements, but in our companion paper \citep{2020arXiv200610156T}, we fit to real observations using the full covariance matrix.

\subsection{Mock near-IR imaging}
\label{sec:data:mockimaging}

To generate a mock catalogue of the cluster galaxies' K-band magnitudes, we run \textsc{subfind} algorithm \citep{2001MNRAS.328..726S} on the particle distribution from the simulations, to identify individual galaxies. We sum their stellar masses, and convert these to $K$-band luminosity based on the relation presented by \cite{2007A&A...476..137A} for the evolution of stellar mass to light ratio, $\left(M/L_K\right)$, with redshift for a sample of quiescent galaxies, and based on the \cite{1955ApJ...121..161S} initial mass function. The power-law fitting function is defined as 
\begin{equation}
\text{log}_{10}\left(M/L_K\right)=a\,z+b,
\end{equation}
{where the mass $M$ and luminosity $L_K$ are in units of $M_{\sun}$ and $L_{\sun}$, respectively.} The best-fit value for parameters $a$ and $b$ from \cite{2007A&A...476..137A} are
\begin{equation}
a = -0.18 \pm 0.04, \quad  b = +0.07 \pm 0.04.
\end{equation}

\subsection{Mock weak-lensing shears}
\label{sec:data:mockshear}

To generate mock weak-lensing observations, we convert the mass distributions into reduced shear.
For the case with projected LSS, we sum the effective convergence from the cluster (section~\ref{sec:data:clusters}) and a random realisation of projected LSS (section~\ref{sec:data:lss}).
Since both convergence $\kappa(\boldsymbol{R})$ and shear $\boldsymbol\gamma(\boldsymbol{R})$ fields are linear combinations of second derivatives of $\varphi(\boldsymbol{R})$, it is possible to directly convert between their Fourier transforms $\hat{\kappa}(\boldsymbol{k})$ and $\hat{\boldsymbol\gamma}(\boldsymbol{k})$
\begin{equation}
\hat{\gamma_1}(\boldsymbol{k}) = \frac{k_1^2 - k_2^2}{k_1^2 + k_2^2}\, \hat{\kappa}(\boldsymbol{k})
\label{gamma1_kappa}
\end{equation}
\begin{equation}
\hat{\gamma_2}(\boldsymbol{k}) = \frac{2 k_1 k_2}{k_1^2 + k_2^2}\, \hat{\kappa}(\boldsymbol{k})\,,
\label{gamma2_kappa}
\end{equation}where $\boldsymbol{k} = (k_1, k_2)$ is the wave vector conjugate to $\boldsymbol{R}$ 
\citep[][hereafter KS93]{1993ApJ...404..441K}. 
To implement this in practice, we pixellate the fields within a 34\arcmin$\times$34\arcmin$\ (2048\times2048$ pixel) grid, add zero padding to twice that linear size to mitigate boundary effects, then use discrete Fourier transforms.
We finally use eq.~\ref{eq:reducedshear} to convert shear $\boldsymbol\gamma(\boldsymbol{R})$ into reduced shear $\boldsymbol{g}(\boldsymbol{R})$.

We generate a mock shear catalogue by randomly placing source galaxies throughout the high-resolution pixellated shear field.
Mimicking typical single-orbit depth {\it HST} observations, we sample $50$\,arcmin$^{-2}$ source galaxies. 
Note that we achieve a uniform density of background galaxies; in real observations, the number density of background galaxies is both clustered, and dips near the centre of a cluster because of obscuration by, and confusion with, its member galaxies. 
To each shear value, we add Gaussian random noise with width $\sigma_{\gamma}=0.36$, representing each galaxy's unknown intrinsic shape, plus uncertainty in shape measurement. This value matches that measured in \emph{HST} measurements near MS\,0451-03 \citep{2020arXiv200610156T}, and is consistent with that measured for faint galaxies in the \emph{HST} COSMOS field \citep[see figure 17 in][]{2007ApJS..172..219L}. It is slightly larger than the intrinsic shape noise referenced elsewhere, because it also includes measurement noise.

\section{Methods} 
\label{sec:method}

In this section, we describe several methods that have been used (or suggested) to analyse the distribution of mass in clusters.
A common theme will be the suppression of noise --- the two main sources of which are projected LSS, and galaxies' intrinsic shapes.
In particular, sophisticated nonlinear noise-suppression techniques have been developed to map the 2D distribution of mass.
Even for measurements that could be obtained directly from the shear field, it may therefore be efficient to first infer (and suppress noise in) a mass map, then to measure equivalent quantities from that.

\subsection{Mass mapping}
\label{sec:method:massrec}

We start by exploring two frequently used methods to reconstruct the distribution of lensing mass: one frequentist, the second Bayesian. 
Where relevant, we adopt parameters in the methods that are typically used by their protagonists.

\subsubsection{Direct inversion with \textsc{KS93+MRlens}}
\label{sec:method:massrec_ks93}

Under the weak-lensing approximation $\boldsymbol{g}=\boldsymbol\gamma$, the KS93 Fourier space relation (see Sect.~\ref{sec:data:mockshear}) can also be used to convert $\boldsymbol\gamma(\boldsymbol R)$ into
\begin{equation}
\hat\kappa(\boldsymbol{k})=
\frac{1}{2}\left(\frac{k_1^2-k_2^2}{k_1^2+k_2^2}\right)\hat\gamma_1(\boldsymbol{k})+
\frac{1}{2}\left(\frac{k_1k_2}{k_1^2+k_2^2}\right)\hat\gamma_2(\boldsymbol{k})\, .
\label{eq:ks93}
\end{equation}
This is a non-local mapping. In observations of the real Universe, any missing shear values (e.g.\ outside the survey boundary or behind bright stars) must be replaced via `inpainting' \citep{2009MNRAS.395.1265P,2019JCAP...11..037R} to avoid suppressing the convergence signal inferred nearby.
We avoid this effect by a using mock shear catalogue that is contiguous and covers a larger area ($34\arcmin\times 34\arcmin$) than the mosaicked {\it HST} imaging of MS\,0451-03. 
We bin the shear field $\boldsymbol{\gamma}(\boldsymbol{R})$ into $0.4\arcmin$ pixels, add zero padding out to 105\arcmin$\times$105\arcmin \citep{Merten:2008qf,Umetsu:2015hda}, and implement eq.~\eqref{eq:ks93} using discrete Fourier transforms.

Noise was suppressed in early incarnations of KS93 by convolving the mass distribution with a larger smoothing kernel whilst in Fourier space. 
We omit this step, and instead filter the final convergence map using the Multi-Resolution method for gravitational Lensing \citep[\textsc{MRLens};][]{2006A&A...451.1139S}. 
This decomposes an image into multiscale starlet wavelets, and applies non-linear regularisation on each wavelet scale.
It aims to retain statistically significant signal but suppress noise through an approach that, under the assumption of a multiscale entropy prior, optimises the False Discovery Ratio (FDR) of false detections to true detections.
\cite{2006A&A...451.1139S} show that \textsc{MRLens} outperforms Gaussian or Wiener filtering at this task, and \cite{2010AIPC.1241.1118P} demonstrate specifically that it improves the reconstruction of non-Gaussian structures like the distribution of mass in galaxy clusters. 
The software implementation\footnote{We implement \textsc{MRLens} using the 2017 June 26 version of software available from \url{https://www.cosmostat.org/software/mrlens}. Note that a 3D extension of this method has also been developed, known as GLIMPSE \citep{2015MNRAS.449.1146L}.} has various free parameters: we use ten iterations during the filtering process, and decompose the noisy 2D convergence map into six wavelet scales, starting at $j=3$. These have size $\vartheta=2^j\,\mathrm{pixels}$. For a starlet wavelet \citep[eq~(11) of][]{2012MNRAS.423.3405L}, the $j=3$ (highest resolution) wavelet is a Mexican hat with full width at half maximum (FWHM) of $0.5\arcmin$.
For comparison to older analyses, we also repeat the analysis after smoothing and rebinning the shear field into larger, $1\arcmin$ pixels. 

\subsubsection{Forward fitting with \textsc{Lenstool}}
\label{sec:method:massrec_lenstool}

We also use \textsc{Lenstool}\footnote{We implement {\sc Lenstool} using version~7.1 of the software available from \url{https://projets.lam.fr/projects/Lenstool/wiki}.} \citep{2009MNRAS.395.1319J} to fit the reduced shear catalogues $\boldsymbol{g}(\boldsymbol{R})$ with a sum of analytic mass distributions.
The field of view considered is the same size as the mosaicked {\it HST} imaging around MS\,0451-03. 
\cite{2009MNRAS.395.1319J} advocate a mass model built of three components.
\begin{itemize}
\item 
\textbf{Cluster-scale halo:}
For clusters that produce strong gravitational lensing, the observed positions of multiple images are typically used to pre-fit the smooth, large-scale distribution of mass \citep{1996ApJ...471..643K,2005MNRAS.359..417S,2011MNRAS.414L..31R,Jauzac:2014xwa}. Like many clusters, our mock data do not include strong-lensing, so we omit this component. Note that our performance forecasts will therefore be conservative, because this information efficiently captures the broad features of a mass distribution in only a few parameters, and removes degeneracies between the remaining parameters that we shall fit \citep{2015MNRAS.446.4132J}.

\item
\textbf{Cluster member galaxies:} 
We model the total mass of each galaxy in the cluster as a tPIEMD (Eq.~\ref{eq:PIMED}). 
Following \cite{2012MNRAS.426.3369J}, 
their core radii, truncation radii and velocity dispersions 
are scaled using empirical relations
\begin{equation}
r_{\text{c}}=r_{\text{c}}^*\left(\frac{L}{L^*}\right)^{\frac{1}{2}}, ~~ \\
r_{\text{t}}=r_{\text{t}}^*\left(\frac{L}{L^*}\right)^{\frac{1}{2}}, \\
\sigma = \sigma^*\left(\frac{L}{L^*}\right)^{\frac{1}{4}}, 
\end{equation}
where $r_{\text{c}}=0.15\text{kpc}$, $r_{\text{t}}=58\text{kpc}$ and $\sigma^*=163.10\text{kms}^{-1}$ for a typical galaxy with $K$-band magnitude $m^*=18.699$ at $z=0.55$. 
These scaling relations describe early-type cluster galaxies \citep{2004ApJ...605..677W}, and assume a constant mass-to-light ratio for all cluster members.

\item
\textbf{Multi-scale, free-form grid:} 
We add a free-form (pixellated) mass distribution with spatially-varying resolution that is adapted to the cluster's light distribution. 
Following \citet[][figure~1]{2009MNRAS.395.1319J}, we initialise a grid of points by drawing a large hexagon over the entire field of view, split into six equilateral triangles with side length $= 1152\arcsec$. If a single pixel inside any of these triangles exceeds a predefined light-surface-density threshold, we split that triangle into four smaller triangles. This refinement continues for six levels of recursion, until the brightest parts of the cluster are covered by the highest resolution grid with $r_c=18\arcsec$. 
We extend this grid into the cluster centre, which is inevitably modelled at the highest resolution.
At the centre of every triangle, we place a 
circular ($q=1$) tPIEMD (Eq.~\ref{eq:PIMED}), with core radius $r_{\mathrm{c}}$ set to the side length of the triangle, truncation radius $r_{\mathrm{t}}=3r_{\mathrm{c}}$, and velocity dispersion that is free to vary. This process represents a prior that light-traces-where-mass-is, rather than explicitly light-traces-mass.
\end{itemize}

We optimise free parameters in this model using the \textsc{MassInf} Markov Chain Monte Carlo algorithm. The parameter space is highly dimensional, so to optimise the multiscale grid, we adopt 
the Gibbs approach \citep[]{2007NJPh....9..447J}, whereby the most discrepant
masses are adjusted during each step of the Markov Chain and as a prior, the initial number of RBFs to explore is set to be $2\%$ \citep[][]{2012MNRAS.426.3369J, 2014MNRAS.437.3969J}.
We apply a prior that the masses are all positive. This need not necessarily be true, since we are really fitting departures from the mean density of the Universe; for example, the convergence of the LSS is consistent with fluctuations around zero (Sect.~\ref{sec:data:lss}). However, the prior is frequently used, and reasonable near a galaxy cluster.
We then finally compute the marginalised mean convergence, and its $68$\% confidence limits.

\subsection{Radial density profiles}
\label{sec:method:radial}

Most analyses of galaxy clusters involve fitting models of an azimuthally-averaged density profile. 
Measuring density profiles is a key test of cosmological structure formation \citep[e.g.\ the `splashback' feature reveals a characteristic build-up of accreted mass, pausing at first apocentre after first core passage][]{Diemer:2014xya} and the nature of dark matter \citep{2013ApJ...765...24N,2015ApJ...814...26N,2019MNRAS.488.3646R}.
Because almost all clusters have irregular features, and approximately half are significantly unrelaxed \citep{2010MNRAS.409..169S}, it is necessary to statistically combine the profiles of many clusters.
This can be achieved by rescaling and averaging their density profiles in radial bins, or by fitting parametric models with radial (or elliptical) symmetry, then averaging the best-fit parameters.

We calculate the radial density profiles of each simulated cluster by azimuthally averaging the reconstructed density maps within linearly spaced annuli of fixed width $\Delta R=25\arcsec$. 
For \textsc{Lenstool} reconstructions, we quote the statistical uncertainty in each annulus, $\sigma_{\rm{stat}}$, determined during the MCMC sampling.
When the signal from projected LSS is included, we add $\sigma_{\rm{LSS}}$, as detailed in Sect.~\ref{sec:data:lss}, such that the total uncertainty error on the density profile, $\sigma_{\rm{tot}}^2=\sigma_{\rm{stat}}^2+\sigma_{\rm{LSS}}^2$.

\subsection{Halo Shapes}
\label{sec:method:shapes}

On large scales, the accretion of matter from
the surrounding large-scale environment plays a key role
in determining the shape and orientation of cluster dark matter halos \citep{Shaw:2005dy}. Halos are not necessarily self-similar \citep[concentric ellipsoids with the same orientation and ellipticity;][]{Suto:2016zqb}, but align with the infall direction of subhalos and surrounding filaments at large radii. Thus, the shape of galaxy clusters is a fundamental probe of the history of its mass accretion.
Numerical simulations with collisionless dark matter predict cluster halos to be triaxial \citep{1992ApJ...399..405W,Jing:2002np}. 
Allowing DM particles to self-interact isotropizes the orbits of dark matter particles, and makes the inner mass distribution more spherical. For a cross-section of $1\,\rm{cm}^2/\rm{g}$, the median minor-to-major axis ratio 100 kpc from the halo centre is $\sim$0.8, compared with $\sim$0.5 with CDM \citep{2019MNRAS.488.3646R}.

We fit an elliptical NFW mass distribution (eq.~\ref{eq:NFW}) to the 2D convergence maps reconstructed from KS93+\textsc{MRLens} or \textsc{Lenstool}, with no noise, with shape noise, with LSS noise or both.
The fit\footnote{We use the \textsc{scipy.minimize} implementation of the \textsc{L-BFGS-B} algorithm \citep{53712fe04a3448cfb8598b14afab59b3}, available from \url{https://docs.scipy.org/doc/scipy/reference/generated/scipy.optimize.minimize.html}.} 
minimize the sum of the squared difference between the reconstructed surface mass density of each BAHAMAS simulated cluster and an elliptical NFW model,
 within a circle of radius $R_\mathrm{ap}$.
We then vary $R_\mathrm{ap}$, to investigate changes between the cluster's inner and outer halos.
During the fits, we fix the centre of the NFW (to the location of the most bound particle) because it is degenerate with axis ratio.
We adopt flat priors on other free parameters: $0.1\leq M_{200}\;(10^{15}\msun)\leq5$, $0.1\leq c_{200}\leq 8$, $0\leq\phi\leq180 $ and $0.1\leq q\leq0.9$, and neglect covariance between adjacent pixels.
The uncertainties of $q$ in this test can be under-estimated. However, it match those in observational data, as we add only one, fixed realisation of LSS along the line-of-sight associated with each cluster.

\subsection{Searches for filaments}
\label{sec:method:filaments}

\begin{figure*}
\centering
\includegraphics[width=0.96\textwidth,trim={0 10mm 0 10mm 0},clip]{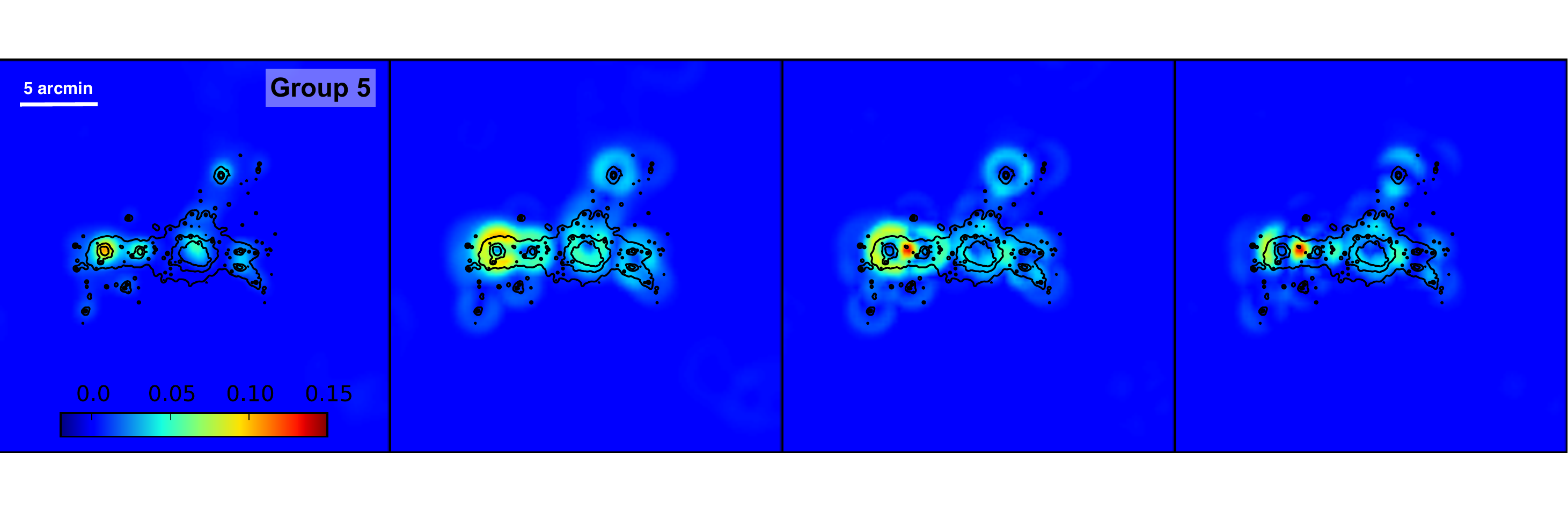}
\caption{An example of aperture multipole moments of various orders, which pick out different features of the noise-free mass distribution of one simulated cluster (Cluster 5, which happens to have several features in the plane of the sky). Moments are calculated after subtracting the large-scale smooth mass distribution. From left to right, panels show: (a) monopole, (b) dipole, (c) quadrupole moments and (d) the radial component of the quadrupole moment. For reference, black contours show the true mass distribution.}
\label{fig:Group1_allmoments}
\end{figure*}

\begin{figure*}
\centering
\includegraphics[width=1.0\textwidth,trim={0 3mm 0 3mm 0},clip]{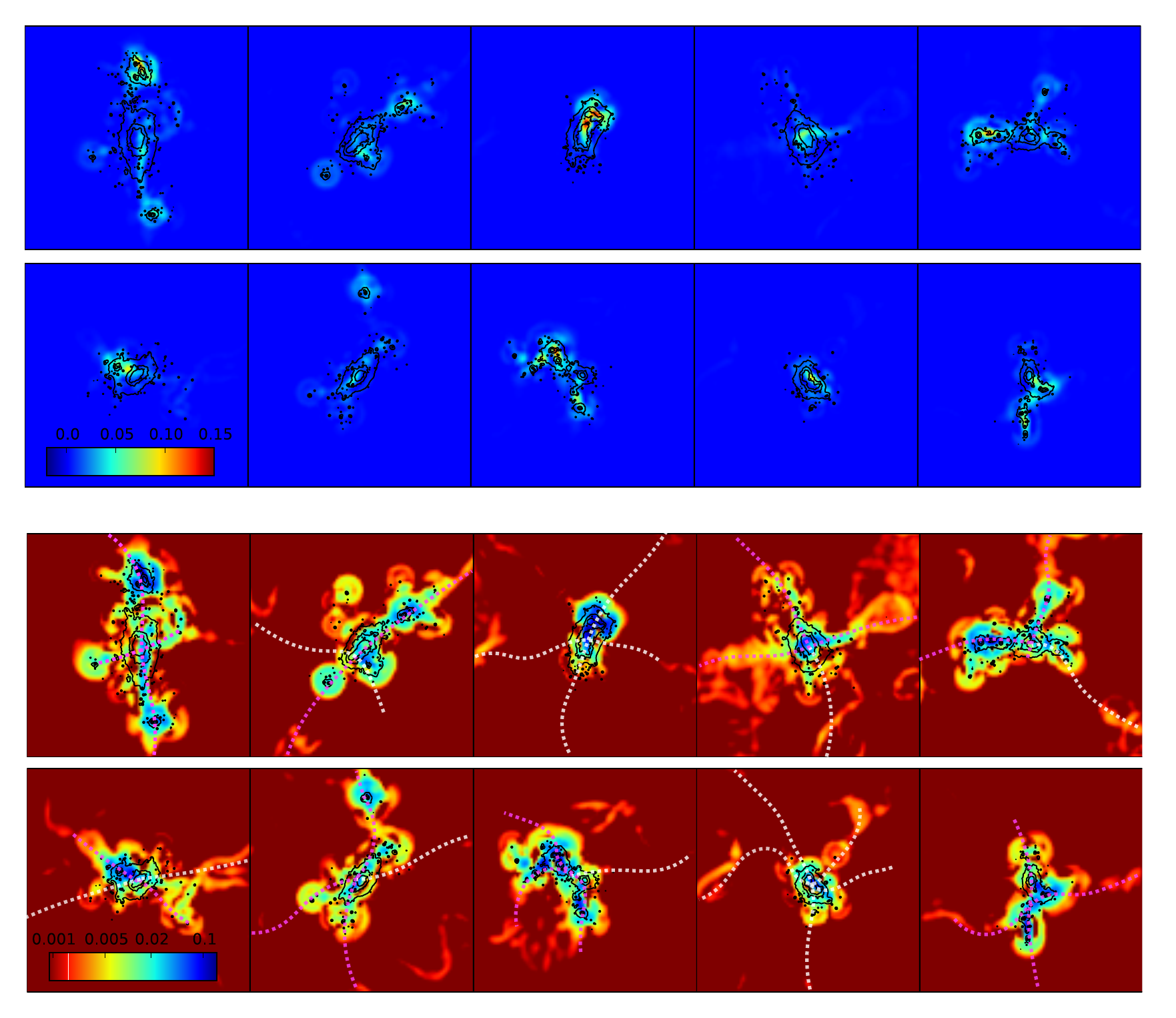}
\caption{
    A combination of aperture multipole moments, $Q$ (equations~\ref{eq:Q_define}--\ref{eq:apsizes}), can be used to identify filamentary features in a mass map.
    Colours (Top panel: linear scale, Bottom panel: logarithmic scale) show $Q$ calculated from the true convergence map (without shape noise or LSS noise; black contours), after subtracting its best-fitting smooth component.
    Dotted lines reproduce the 40 filaments from figure~\ref{fig:truekappa}. 
    The 22 filaments successfully identified using $Q$ and the procedure described in section~\ref{sec:method:filamentid} are highlighted in magenta.
\label{fig:Qmeasurements_true}}
\end{figure*}

Dark matter and gas are accreted onto a cluster mainly through filaments that connect it to the `cosmic web'.
Filaments are key transition regions in the evolution of galaxy morphology \citep{Pandey:2006rj,2007A&A...464..815E,2014MNRAS.445..988N,2017A&A...600L...6K,Liu:2019mfb,Martizzi:2019sax} and star formation \citep{2009MNRAS.399.1773C,2010MNRAS.408.1818W,2015MNRAS.451.3249A,2016MNRAS.457.2287A,2019MNRAS.487.1315Y}.

Filaments are much lower density environments than a cluster, so appear in gravitational lensing observations with correspondingly lower signal-to-noise.
While it is possible to search for filaments directly in shear data \citep{Dietrich:2004kf,2012Natur.487..202D,2012MNRAS.426.3369J}, we explore whether it is efficient to leverage the de-noising techniques developed for mass mapping, then to analyse the inferred convergence field.

\subsubsection{Removing the smooth mass component}

First, we subtract the smooth distribution of mass in the clusters, which would otherwise dominate the lower density contrast in the filaments.

We fit mock reduced shear data
(with or without LSS and galaxy shape noise), using an elliptical NFW potential. This model has 6 free parameters: the coordinates of the centre of mass, ($x_\mathrm{c}$, $y_\mathrm{c}$), the ellipticity, $e=(1-q^2)/(1+q^2)$ where $q$ is the axis ratio, the position angle, $\phi$, the scale radius, $r_s$, and the concentration, $c$. 
We set flat priors on $x_\mathrm{c}$ and $y_\mathrm{c}$ within a $15\arcsec\times15\arcsec$ box centred on the most bound particle, 
and flat priors on 
$e\in[0.05,0.7]$, $\phi\in[0,180]$, $r_s\in[50,1000]$\,kpc, and $c\in[0.5,10]$. 
Note that we introduce ellipticity to this model via a coordinate transformation to the gravitational potential (rather than the mass, as in Sect.~\ref{sec:theory:analytic}) because code to achieve this already exists within \textsc{Lenstool}\footnote{An elliptical gravitational potential produces a `boxy' mass distribution if $e>0.6$. However, for the low values of ellipticity that we obtain, the maximum distance $\delta R$ between a projected density contour and a true ellipse is $\delta R/R<10\%$ \citep[see figure 6 in][]{2002A&A...390..821G}.}. The smooth distribution of mass in most simulated clusters is well approximated by a single potential. However, we use two to fit bimodal clusters 1, 2 and 9, and three for cluster 3.

We then subtract the best-fit smooth halos from the convergence maps.  
Since the mass distribution of simulated clusters cannot be perfectly described by elliptical NFW potentials, 
small residuals are left near the cluster centre.
Such residuals do not impact searches for filaments at much larger radii.

\subsubsection{Aperture multipole moments}

\cite{Schneider:1996yy} first suggested looking for substructures or filaments using multipole moments of a convergence field within circular apertures. These are 
\begin{equation}
    Q_n(\boldsymbol{R})=\int_0^{\infty} |\boldsymbol{R}'-\boldsymbol{R}|^n\, \mathrm{e}^{ni\phi}~U_n(|\boldsymbol{R}'-\boldsymbol{R}|)~\kappa(\boldsymbol{R}')~\mathrm{d}^2\boldsymbol{R}'\,,
    \label{eq:multipoleint}
\end{equation}
where $n$ is the order of the multipole, ($R$, $\phi$) are polar coordinates, and $U_n(R)$ is a radially symmetric weight function, for which \cite{Dietrich:2004kf} suggested
\begin{equation}
U_n(R)=
\begin{cases}
1-\left(\frac{R}{R_{\mathrm{max},n}}\right)^2 & \text{for $R\leqslant$ $R_{\mathrm{max},n}$}\\
0 & \text{otherwise}.
\end{cases}
\end{equation}
Eq.~\eqref{eq:multipoleint} can also be  expressed in terms of shear measurements, which \cite{Dietrich:2004kf} used to detect filament candidates in close pairs of clusters. Since modern mass reconstruction methods successfully suppress noise, we attempt instead to measure multiple moments directly from the pixellated convergence field
\begin{equation}
Q_n(\boldsymbol{R})=A_{\rm{pix}}\sum_{i=1}^{N_\mathrm{pix}}\,R_{i}^{n}~\mathrm{e}^{ni\phi_i}~U_n(R_i)~\kappa(\boldsymbol{R_i})\, ,
\label{eq:k_moments}
\end{equation}
where $N_\mathrm{pix}$ is the total number of pixels inside the aperture and $A_{\rm{pix}}$ is an area per pixel.
For $n>0$, $Q_n$ is complex; we shall generally take its modulus, $|Q_n|$.

Multipoles of different orders highlight different features in a mass distribution (see figure~\ref{fig:Group1_allmoments}). Monopole moments ($n=0$) are the aperture mass or normalisation. Dipole moments ($n=1$) are the local gradient of a convergence field. They form ring-like structures around mass clumps. Quadrupole moments ($n=2$) are the locally-weighted curvature or Hessian of the convergence field. As \cite{Dietrich:2004kf} explain using a toy model, linear overdensities with a lower mass on either side (i.e.\ filaments) have large quadrupole moments.
However, regions {\it between} two substructures also have large quadrupole moments. 
To identify the former and suppress the latter, 
\cite{2010MNRAS.401.2257M} suggested combining multipole moments
\begin{equation}
Q\equiv\alpha_0\,Q_0+\alpha_1\,Q_+1+\alpha_2\,Q_2+...
\label{eq:Q_define}
\end{equation}
where the constants, $\alpha_i$, can be adjusted to 
boost a signal of interest. 
We have tried different combinations and aperture sizes, and find that a choice of
\begin{eqnarray}
 \label{eq:apcoeffs}
 \alpha_0=-\alpha_1=0.7 \quad \mathrm{and} \quad \alpha_2=1\,, ~~~~~~~~~~~~~~~~~\\
 \label{eq:apsizes}
 R_{\mathrm{max},0}=1\arcmin \quad \mathrm{and} \quad R_{\mathrm{max},1}=R_{\mathrm{max},2}=2\arcmin\,.
\end{eqnarray}
typically highlights narrow filaments (see figure~\ref{fig:Qmeasurements_true}).
The quadrupole term is sensitive to linearly extended mass distributions, and the rings that it adds around substructures are removed by the negative dipole term. The monopole term fills in the subtracted mass, and suppresses regions between two substructures but without mass.

\subsubsection{Filament identification} 
\label{sec:method:filamentid}

\begin{figure}
    \centering
    \includegraphics[width=\columnwidth]{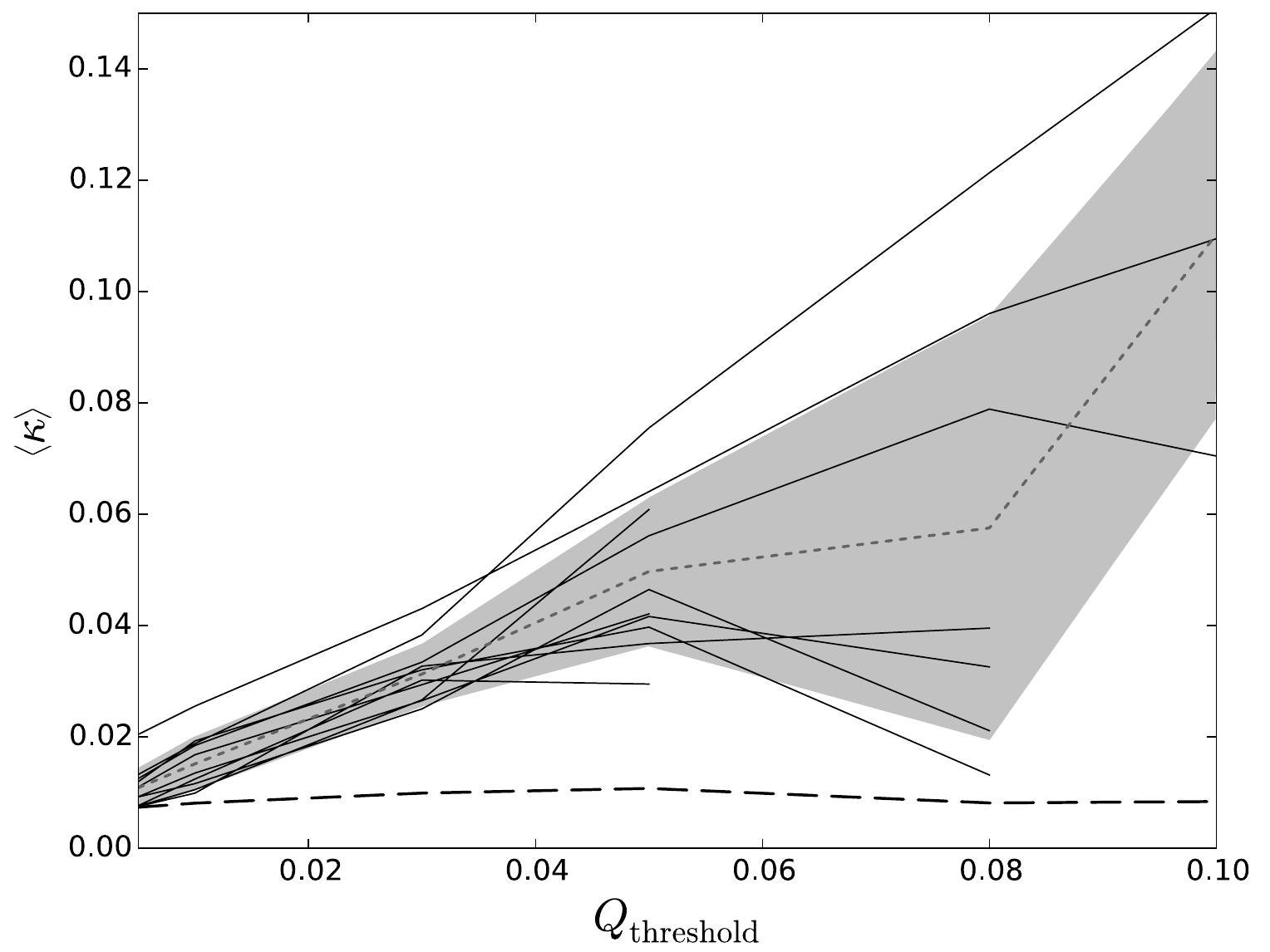}
    \caption{
    A combination of aperture multipole moments, $Q$ (equations~\ref{eq:Q_define}--\ref{eq:apsizes}), can be used to identify features in a mass distribution with filamentary topology (see figure~\ref{fig:Qmeasurements_true}) and higher density than the background.
    Solid lines show the mean projected density $\langle\kappa\rangle$ inside a contour defined by $Q_\mathrm{threshold}$, for all 10 simulated clusters. The dotted line and shaded region show their mean and standard deviation.  
    The normalisation of coefficients \eqref{eq:apcoeffs} is chosen so that  $\langle\kappa\rangle=Q_\mathrm{threshold}$.
    The lower dashed line shows the mean convergence, weighted by the number of pixels that contain $Q>Q_\mathrm{threshold}$.
    \label{fig:multiq_mass}}
\end{figure}

To identify individual filaments, we search for spatially extended regions with $Q$ above a threshold $Q_\mathrm{threshold}$.
The normalisation of coefficients in eq.~\eqref{eq:apcoeffs} conveniently ensures that regions inside a contour $Q_\mathrm{threshold}$ have mean convergence $\langle\kappa\rangle\approx Q_\mathrm{threshold}$ (figure~\ref{fig:multiq_mass}).
We identify as possible filaments any region with $Q>Q_\mathrm{threshold}$ in a contiguous area or multiple peaks with total area $>1.13$\,arcmin$^2$, that is aligned within $\sim45^\circ$ of the radial direction to the cluster centre.
Applied to noise-free data and using $Q_\mathrm{threshold}=0.005$, this recipe identifies 22 of the 40 filaments, all of which are real, i.e.\ $55\%$ {\it completeness} (the number identified divided by the true number) and $100\%$ {\it purity} (the number identified that are true divided by the number identified).
The identified filaments are highlighted in magenta in figure~\ref{fig:Qmeasurements_true}.

\subsubsection{Additional noise suppression strategies} 
\label{sec:method:denoising}

Measurements of multipole moments will be more difficult in noisy data --- especially for high $n$ moments, where the diverging $|\boldsymbol{R}'-\boldsymbol{R}|^n$ term is particularly sensitive to noise in $\kappa$ near the aperture boundary. We shall explore three strategies to reduce noise. 
First, noise can be averaged away by enlarging the aperture. 
However, signal is also averaged away for a filter than is not matched to the size of the feature -- and filaments are relatively narrow, even around clusters at low redshift. 
Second, negative noise peaks can be eliminated by forcing $\kappa=\mathrm{max}\{\kappa,0\}$.
Negative convergence is physically possible, because convergence represents deviation from the mean cosmic density; but it is unlikely along the line-of-sight to even a low density structure, and probably noise rather than signal.
Third, we could assume that all filaments extend radially away from the cluster, while noise is isotropic, and suppress quadrupole and dipole moments whose phases are tangential. We calculate
\begin{equation}
Q_{n,\text{projected}}=|Q_n|\,\cos{(\phi-\theta)}, \quad \mathrm{with\;\;}{n=1,2}
\label{eq:proQ}
\end{equation}
where $\theta$ is an phase angle of $Q_n$.
Figure~\ref{fig:Group1_allmoments}(d) shows the projected quadrupole moments in the noise-free case, as an example.

\begin{table*}
  \centering
  \resizebox{0.95\textwidth}{!}{
  \begin{tabular}{lcccccc}
   \toprule
     & \multicolumn{3}{c|}{$\sigma_\kappa$} & \multicolumn{3}{c}{$\sigma_\kappa^\mathrm{obs}$} \\
     & \textbf{Full mock} & \textbf{Shape noise only} & \multicolumn{1}{c|}{\textbf{LSS noise only}} & \textbf{Full mock} & \textbf{Shape noise only} & \textbf{LSS noise only} \\

    \midrule
    \midrule
    \textbf{KS93} (pixel scale $0.4\arcmin$) & $0.088\pm0.001$ & $0.091\pm0.001$ &$0.017\pm0.002$ & $0.090\pm0.002$ & $0.092\pm0.001$ &$0.027\pm0.006$ \\
     \textbf{KS93} (pixel scale $1\arcmin$) & $0.037\pm 0.001$ & $0.037\pm0.001$ &$ 0.013\pm 0.002$ & $0.042\pm 0.003$ & $0.039\pm0.002$ &$0.024\pm 0.006$ \\
    \midrule
    \textbf{\textsc{KS93+MRLens}}  & $0.026\pm0.001$ & $0.024\pm0.001$ & $0.014\pm0.002$ & $0.032\pm0.004$ & $0.028\pm0.002$ & $0.024\pm0.006$ \\
    ~~~\textbf{\textit{High mass clusters}} & $0.026\pm0.001$ & $0.024\pm0.001$& $0.016\pm0.001$ & $0.035\pm0.004$ & $0.030\pm0.002$ & $0.029\pm0.005$ \\
    ~~~\textbf{\textit{Low mass clusters}} & $0.026\pm0.002$ & $0.024\pm0.001$& $0.012\pm0.001$ & $0.029\pm0.003$ & $0.026\pm0.001$ & $0.019\pm0.003$ \\
     \midrule
      \textbf{\textsc{Lenstool}} & $0.015\pm0.004$ & $0.012\pm0.003$& $0.013\pm0.004$ & $0.023\pm0.007$ & $0.022\pm0.007$ & $0.024\pm0.008$ \\
    ~~~\textbf{\textit{High mass clusters}} &  $0.018\pm 0.002$ & $0.014\pm0.002$& $0.016\pm0.003$ & $0.031\pm0.005$ & $0.030\pm0.005$& $0.030\pm0.008$ \\
    ~~~\textbf{\textit{Low mass clusters}} & $0.012\pm0.002$ & $0.010\pm0.002$& $0.010\pm0.001$ & $0.018\pm0.003$ & $0.018\pm0.003$& $0.019\pm0.004$ \\
    \bottomrule

  \end{tabular}}
  \caption{Noise level in mass maps created using different methods, measured as the standard deviation of all pixels inside a field of view equivalent to {\it HST} observations of MS\,0451-03. Central values and uncertainties show the mean and standard deviation between clusters.
The first three columns show deviations from the true, noise-free mass map; the second three columns show deviations from zero --- which can be compared to observations of the real Universe. 
The 2$^\textrm{nd}$, 3$^\textrm{rd}$, 5$^\textrm{th}$ and 6$^\textrm{th}$ columns refer to analyses in which the shear catalogues contain only certain sources of noise, so their relative effect can be assessed. 
The first two rows quantify the performance of KS93 direct inversion, with noise suppressed only via convolution with a top hat window function.
The middle rows suppress noise using {\sc MRLens}.
The bottom rows use {\sc Lenstool}.}
  \label{tab:sigmak}
\end{table*}

\section{Results \& Discussion}
\label{sec:results}

To the ten simulated clusters presented in Sect.~\ref{sec:data}, we shall now apply the analysis methods described in Sect.~\ref{sec:method}. We compare the reconstructed convergence maps, radial density profiles and halo shapes, to the known, true distribution of mass. We then search for observable signatures of filaments extending from the clusters.
For all these analyses, we quantify the impact of the two main sources of noise in weak lensing measurements: unrelated LSS projected by chance along the line of sight to the cluster (Sect.~\ref{sec:data:lss}), and the intrinsic shapes of background galaxies (Sect.~\ref{sec:data:mockshear}).

\subsection{Mass mapping}

\begin{figure*}
\centering
\includegraphics[width=0.7\textwidth]{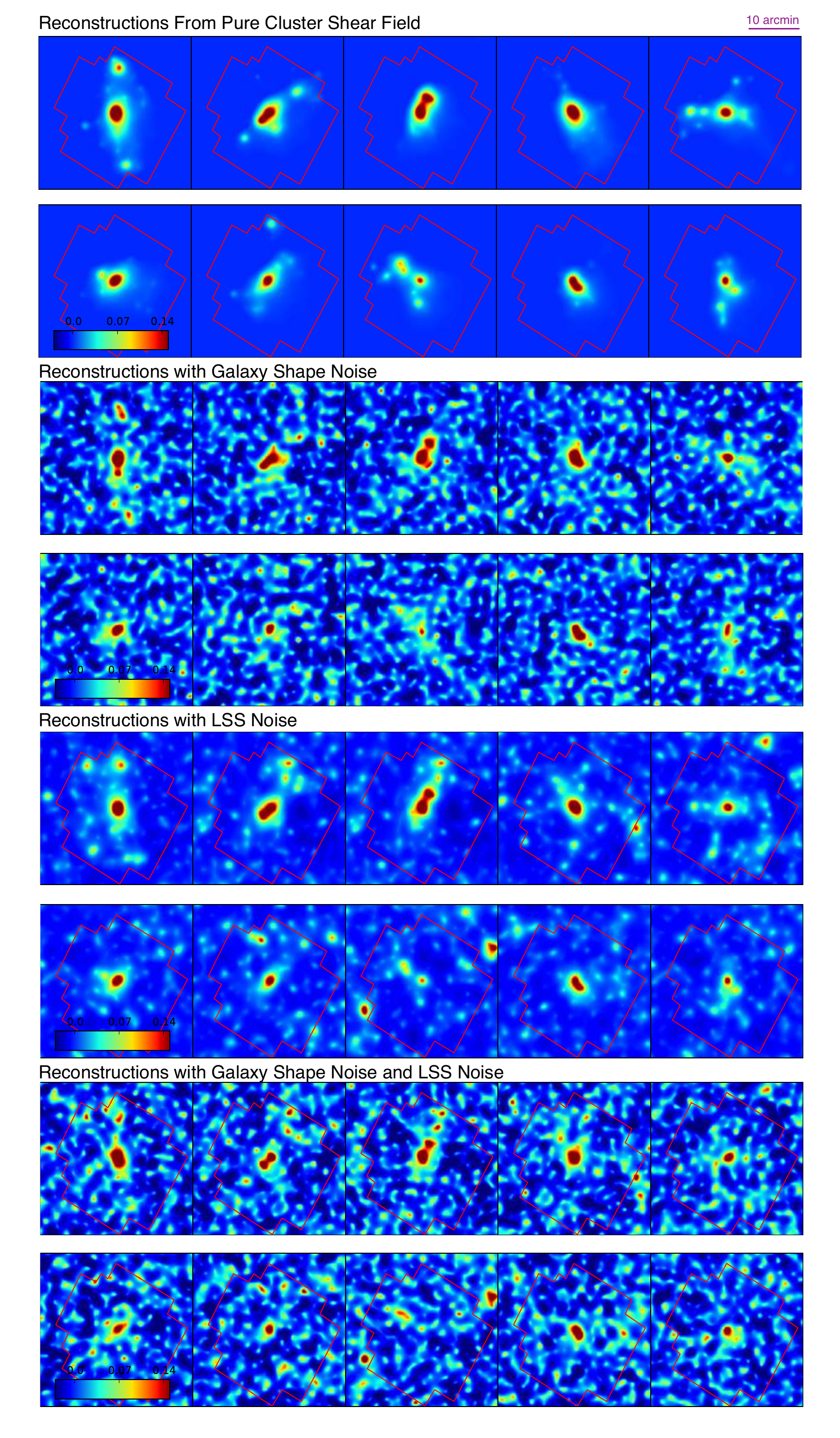}
\caption{Projected mass maps of the 10 simulated clusters reconstructed using the \textsc{KS93+MRLens} direct inversion method, including different components of noise.
{\it Top panels:} reconstruction with no noise.
{\it Second panels:} including only shape noise from 50 background galaxies per square arcminute.
{\it Third panels:} including only projected large-scale structure.
{\it Bottom panels:} including both sources of noise simultaneously.
Colour scales are identical for all panels. 
For reference, red lines indicate the field of view of the largest \emph{HST} mosaic obtained around a massive galaxy cluster, MS\,0451-03. }
\label{fig:mrlensrec}
\end{figure*}

\begin{figure*}
\centering
\includegraphics[width=0.72\textwidth]{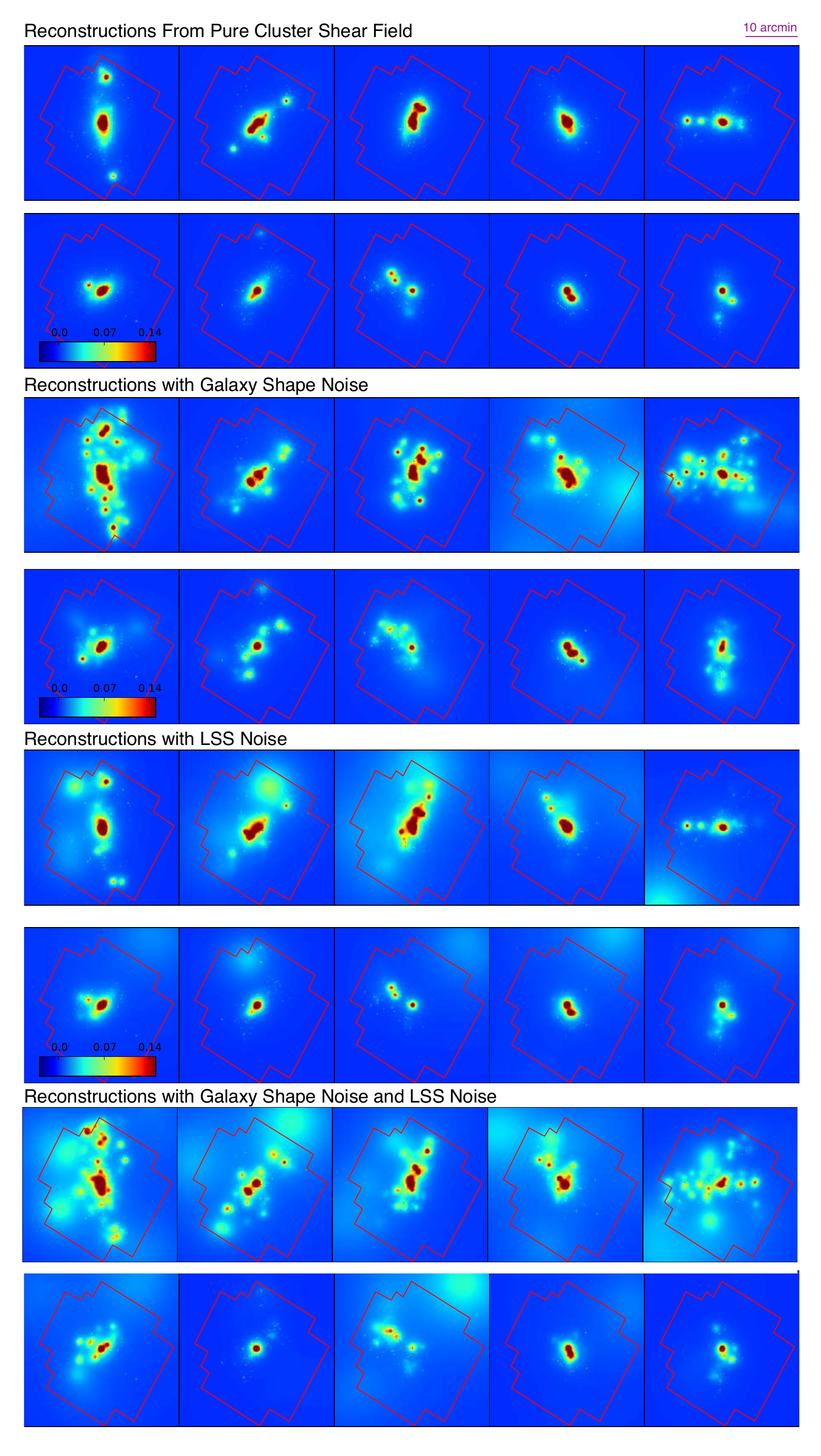}
\caption{Same as figure~\ref{fig:mrlensrec}, but reconstructed using \textsc{Lenstool}.}
\label{fig:Lenstoolrec}
\end{figure*}

We quantify the precision and accuracy of mass maps produced by \textsc{KS93+MRLens} (figure~\ref{fig:mrlensrec}) and \textsc{Lenstool} (figure~\ref{fig:Lenstoolrec}) by comparing them to the noise-free distributions of mass, $\kappa_{\mathrm{true}}$ (which includes only the mass of the cluster, not projected LSS). 
We first measure deviations from this truth, $\kappa_{\rm{res}}\equiv\kappa-\kappa_\mathrm{true}$, to obtain the residual maps. For each map, we compute the noise level $\sigma_\kappa$, defined as the root mean square (rms) deviation from the mean of $\kappa_{\rm{res}}$, over all pixels in a field of view equivalent in size to the \emph{HST} observations of MS\,0451-03. We then average the performance of each method over all 10 clusters (table~\ref{tab:sigmak}).

In observations of the real Universe, $\sigma_\kappa$ cannot be calculated because there is no privileged knowledge of $\kappa_\mathrm{true}$. 
For comparison with observations, we therefore also measure $\sigma_\kappa^\mathrm{obs}$, the rms deviation from the mean of $\kappa$. 
We find values of $\sigma_\kappa^\mathrm{obs}$ roughly consistent with $\sigma_\kappa$ being added in quadrature to an irreducible component that is the rms deviation from the mean of $\kappa_\mathrm{true}$,  
$0.022\pm0.0007$ on average (0.027 for the five highest mass clusters, or 0.017 for the five lowest).

\subsubsection{Direct inversion mass reconstruction}
\label{sec:results:mrlens}
  
{\sc MRLens} suppresses galaxy shape noise by a factor 3.8 (a factor 1.5 better than smoothing with $1\arcmin$ pixels, and retaining higher spatial resolution).
However, galaxy shapes still contribute more noise to the mass maps than (physically real) LSS noise.
Spurious noise peaks are found in all regions of the field of view.
Massive substructures with $\kappa>0.096$ can be detected with S/N\,$>3$.

Mass reconstructions using \textsc{KS93+MRLens} are statistically consistent with being unbiased.
Both positive and negative noise fluctuations are produced, at all radii.
The mean residual of maps with both sources of noise is $\langle\kappa_{\rm{res}}\rangle=-0.0005\pm 0.0018$, where the averaging is over 10 clusters, and the uncertainty is the standard deviation between them.
The marginally negative mean may be because density is underestimated in a small region near cluster cores (see Sect.~\ref{sec:results:radialdensity}).

\subsubsection{Forward-fitting mass reconstruction}
\label{sec:results:lenstool}

\textsc{Lenstool} suppresses noise even further. 
Galaxy shape noise is an {additional} factor 2 lower than \textsc{KS93+MRLens} (averaged across the field of view) --- with the similar level as the LSS noise.

The spatial distribution of noise is nonuniform.
A \textsc{Lenstool} reconstruction has more freedom in regions with a high resolution free-form grid (section~\ref{sec:method:massrec_lenstool}), such as the cluster core and associated substructures. 
Spurious $\kappa$ peaks appear preferentially in those regions, even when we replace the shear catalogue with one that contains only (spatially uniform) galaxy shape noise.
To further investigate this effect, we split the ten clusters into two subsamples: higher mass (clusters 1 to 5), and lower mass (clusters 6 to 10).
Multi-scale grids of the high mass sample have larger high-resolution regions, resulting in noisier maps on average. 
Assessing the S/N of any identified peak must therefore involve bootstrap analysis at the specific region of interest.
This confirms \citet{2014MNRAS.437.3969J}'s similar assessment of the 
performance of \textsc{Lenstool}. 
For many scientific purposes, spatially varying noise is a useful feature: the lower resolution and positive definite constraints help to suppress positive LSS noise and remove negative noise at large radii. 
Even filaments contain a statistically significant overdensity of galaxies \citep{2020arXiv200309697G}, so the reconstruction can be given sufficient flexibility to include (rather than suppress) them.

Mass reconstructions using \textsc{Lenstool} slightly overestimate the total mass, because of its positive-definite constraint.
Averaged over the field of view, the mean residual of maps with both sources of noise is $\langle\kappa_\mathrm{res}\rangle=0.0088\pm 0.0064$ (we quote the mean of $\kappa_\mathrm{res}$ for 10 clusters and the standard deviation between them).

\subsection{Radial density profiles}
\label{sec:results:radialdensity}

\begin{figure*}
    \centering
    \includegraphics[width=0.69\textwidth,trim={0 3mm 3mm 3mm 0},clip]{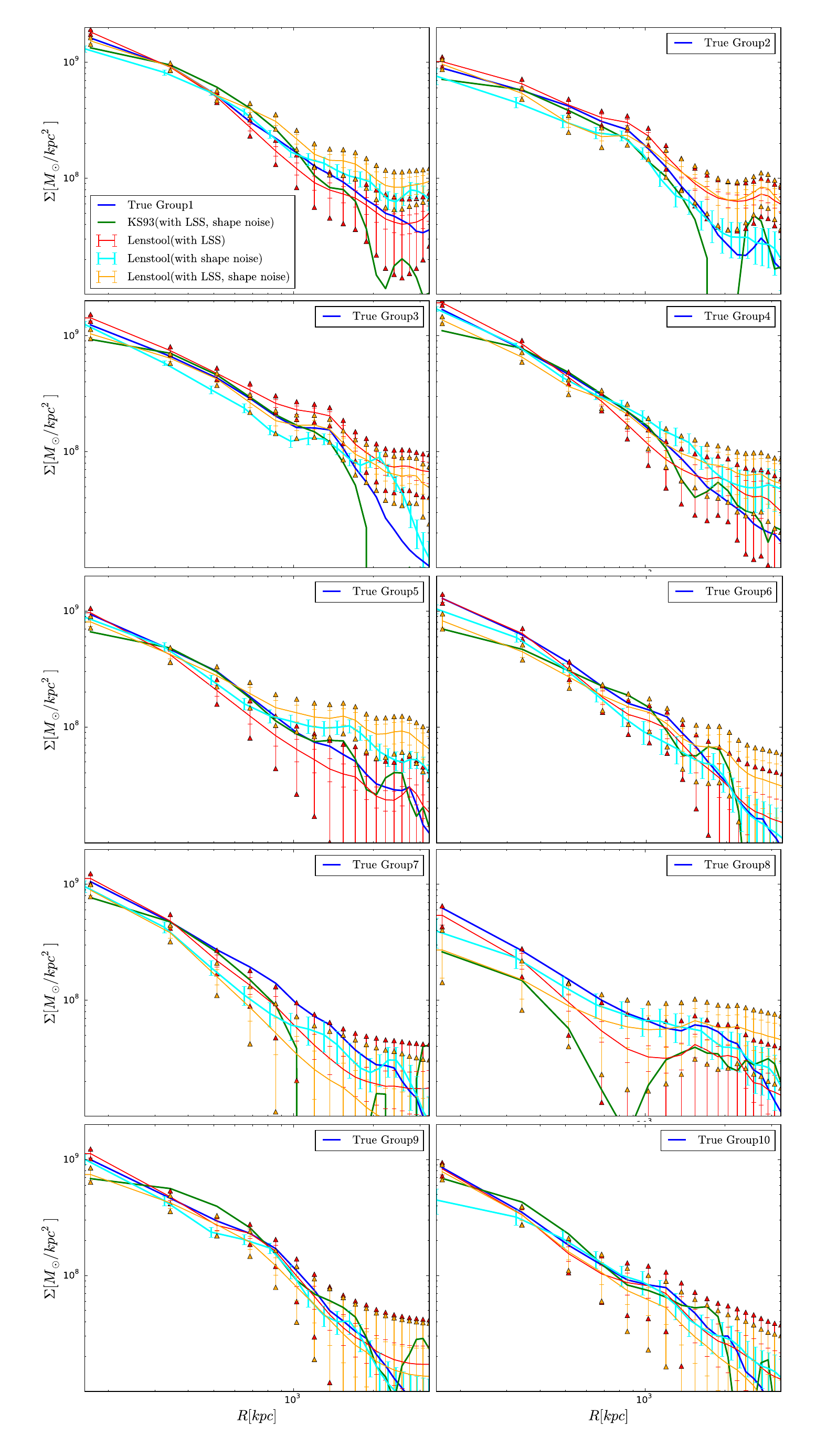}
    \caption{Surface mass density profiles for all 10 simulated clusters. Blue solid lines show the the density profile calculated from the true mass distribution in Fig~\ref{fig:truekappa}. Green solid lines are the density profiles of KS93+\textsc{MRLens} reconstructed maps after adding shapes noise and LSS. Cyan, orange, and red lines show the results recovered by \textsc{Lenstool} including shape noise, projected LSS, and both shape noise and LSS, respectively. Error bars with line caps are statistical errors from the MCMC sample. Error bars with triangle caps are total errors which is the combination of statistical errors with the estimated noise from the projected LSS (eq.~\ref{eq:sigmaLSS}).
    \label{fig:10_profile}}
\end{figure*}

\begin{figure*}
    \centering
    \includegraphics[width=1.0\textwidth]{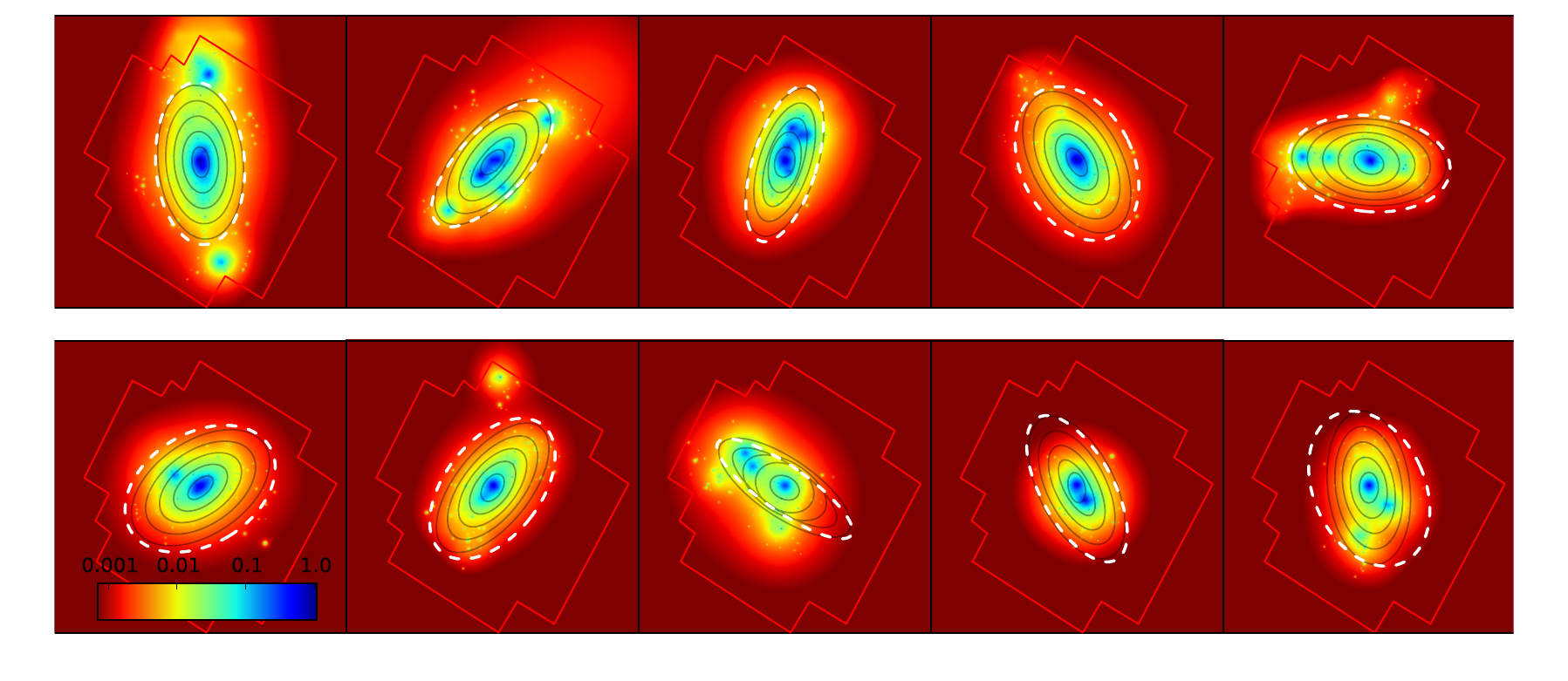}
    \caption{
    Elliptical eNFW models fitted to the {\sc Lenstool} mass maps are $\sim6\%$ too round, on average (see figure~\ref{fig:axisratio}).
    Black ellipses have the same axis ratio of the true mass distribution (see figure~\ref{fig:truekappa}) inside annulus $R<R_\mathrm{ap}$, where different values of $R_\mathrm{ap}$ are indicated by the length of the major axis.
    White dashed ellipses show the axis ratio 
    measured from masked Lenstool reconstructions, inside the largest $35\arcsec<R<R_\mathrm{ap}$.
    The background image shows the mass distribution reconstructed by {\sc Lenstool}, as in figure~\ref{fig:Lenstoolrec} but with a logarithmic scale to highlight one problem with the {\sc Lenstool} method: overly circular central halos.
    }
    \label{fig:logLenstool}
\end{figure*}

We recover the clusters' density profiles by azimuthally averaging the convergence maps (figure~\ref{fig:10_profile}). 
The smoothing inherent to \textsc{KS93+MRLens} results in an underestimation of density in the cluster core, and an overestimate just outside. This biases the inner profile slope that is often used to distinguish between cusps and cores.
\textsc{Lenstool} is accurate in the cluster core, because its basis functions have a density profile that matches those of the simulated clusters. This  is not affected by \textsc{Lenstool}'s positive-definite constraint, because the true mass distribution is very positive near the core.
In the cluster outskirts, \textsc{Lenstool} strongly suppresses galaxy shape noise, and the reconstruction is dominated by LSS noise. 
Because of the positive-definite constraint, this is also potentially biased.
The amplitude of LSS noise varies a great deal depending on environments along the line-of-sight LSS, but we typically find artificial boosts in inferred density of up to $\sigma_{\rm{LSS}}=4\times10^7\,\msun$\,/\,kpc$^2$, at large projected radii, $R>1000$\,kpc.
This effect must be taken into account when measuring properties at large radius (e.g.\ $M_{200}$, $c_{200}$, splashback radius). 
To militate against this, measurements of galaxy redshifts will be invaluable to disentangle structures connected to the cluster from those lying in the foreground or background.

\subsection{Halo shapes} 
\label{sec:results:haloshapes}

\begin{figure}
    \includegraphics[width=0.5\textwidth]{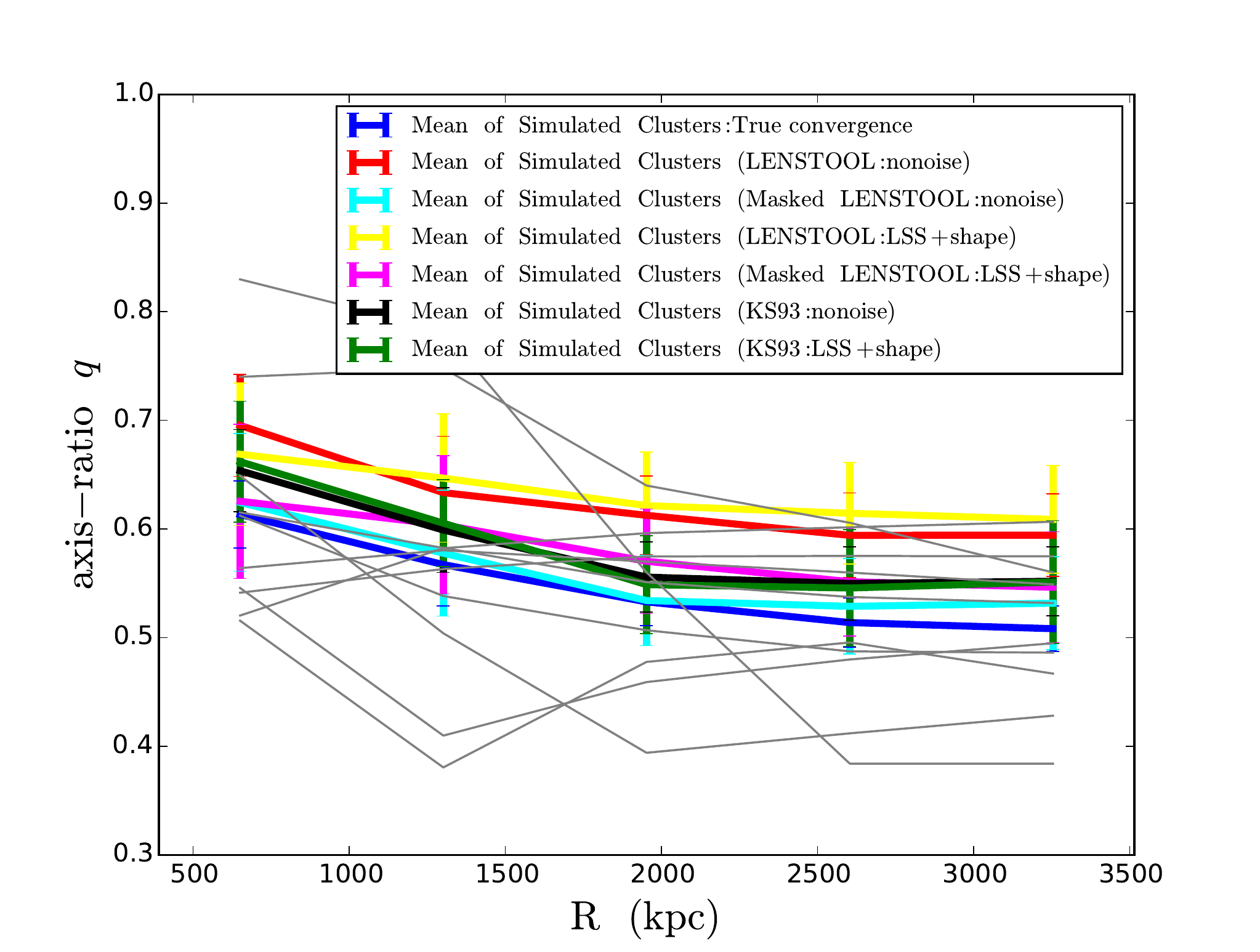}
    \caption{Best-fit axis ratios of the mass distribution in galaxy clusters, as a function of projected, clustercentric radius $R$. Grey lines show the BAHAMAS simulated clusters, whose axis ratio profiles are measured from the true mass distribution. Blue lines show the mean and standard deviations from this set of clusters. Black (green) lines show the mean axis-ratio and its scatter measured from noise-free KS93+\textsc{MRLens} reconstruction (with LSS and shape noise). Red (yellow) lines show the mean results measured from noise-free \textsc{Lenstool} reconstruction (with LSS and shape noise). Cyan (magenta) lines show the axis-ratio measured from the masked $R<35\arcsec\,(228$ kpc) \textsc{Lenstool} reconstruction (with LSS and shape noise).}
\label{fig:axisratio}
\end{figure}

Both mass reconstruction methods produce distributions that are rounder than the truth (figure~\ref{fig:axisratio}).
eNFW models fitted to the reconstructed mass maps (figure~\ref{fig:mrlensrec},~\ref{fig:Lenstoolrec}) have a higher mean axis ratio $\langle q\rangle$ than models fitted to the true mass maps (figure~\ref{fig:truekappa}).
However, they successfully capture the decrease in $\langle q\rangle(R)$ at large radii that is seen in the true mass maps \citep[reflecting a transition from dominant baryonic effects to the infall of structures along filaments;][]{Suto:2016hoj}.
The orientation of most inner ($R=650$\,kpc) and outer ($R=3$\,Mpc) halos also remain aligned within $\Delta\phi\leq10^{\circ}$, matching the true distributions \citep[and also the simulations by][]{Despali:2016pkj}.
Two exceptions to this are clusters 5 and 9, which have complex cores and $\Delta\phi=17^{\circ}$ and $\Delta\phi=15^{\circ}$. 
This likely indicates a transitory state during a major merger. 

Using \textsc{KS93+MRLens} leads to inferred values of $\langle q\rangle$ that are too high by about $6\%$.
The level of bias is not significantly influenced by either source of noise in the shear catalogue (although adding noise increases scatter in individual measurements of $q$ as expected).
It is likely due to the isotropic blurring associated with pixellisation and \textsc{MRLens} filtering.

Using \textsc{Lenstool} leads to inferred values of $\langle q\rangle$ that are too high by $10\%$ in the cluster core and $15\%$ in the outskirts.
The bias appears to be caused by two effects:
\begin{itemize}
\item
The mass distribution is built from components that are all individually spherical. 
If the dominant halo in the cluster core is anomalously spherical (see clusters 4, 5, 8 or 10 in figure~\ref{fig:logLenstool}), it can bias the apparent axis ratio of the mass inside a circle by up to 10\%, almost regardless of the size $R_\mathrm{ap}$ of that circle.
Substructures far from the centre of the cluster look surprisingly uniform, but this does not affect measurements of the overall shape.
\item
The mass distribution is constrained to be positive definite. 
In the absence of noise, this has no effect.
If we add galaxy shape noise, it is also relevant that the reconstructed mass distribution is higher resolution (has more freedom) along its major axis.
The positive-definite bias in noise artefacts then exaggerates the major axis, reducing $\langle q\rangle$ by $\sim$$5\%$.
If we add LSS noise, $\langle q\rangle$ increases by $8\%$ because there is a larger area at close to zero convergence along the minor axis. 
\end{itemize}
It is possible to mitigate the first effect by masking the cluster core. 
We successfully recover the true axis ratio when fitting an eNFW using to {\em noise-free} data inside an annulus $35\arcsec<R<R_\mathrm{ap}$ (instead of a circle of radius $R_\mathrm{ap}$).
Fitting inside annuli also decorrelates measurements of $\langle q\rangle$ at different radii, and steepens the apparent gradient in $\langle q\rangle(R)$. 
Note that the second effect still increases $\langle q\rangle$ by $\sim$$6\%$ in the presence of both sources of noise.

A different strategy to mitigate sphericity bias could be to pre-fit the axis ratio of central halos, then hold them fixed while the rest of the grid is constrained. 
A similar two-step process happens naturally in most combined analyses of strong plus weak lensing, where strong lensing information constrains a cluster core. 
This bias should therefore not affect {\sc Lenstool} strong lensing analyses.
However, it would be difficult to characterise statistical uncertainty in such analysis, because shear data would be used twice.

\subsubsection{Comparison with previous studies}

Previous work by simulators to measure the shape of cluster-scale halos split into two distinct conclusions.
\cite{Hopkins_2005}
found that 2D cluster ellipticity increases with clustercentric radius, in agreement with our results. However, they also found that the ellipticity is $\epsilon\approx0.05 z + 0.33$ for the redshift range $0<z<3$, which implies $q=0.64$ at the $z=0.55$ redshift of our simulated clusters. 
Similarly, \cite{Ho:2005ea} 
found $q\sim0.616$ for halos with masses $M>10^{14}\msun$ at $z=0.55$ assuming $\Omega_m=0.3$, and $\sigma_8=0.7$, and little dependence upon cosmological model. 
Both of these results are slightly rounder than our measurement of $\langle q\rangle_{\rm{true}}\sim0.55\pm0.03$. 

More recently, \cite{Despali:2016pkj} found that $M\sim10^{15}\msun/h$ halos in the SBARBINE N-body simulations had more elliptical shapes, with $q\sim 0.55$. 
\cite{Suto:2016zqb} studied  
the probability distribution function (PDF) of $q$ from projected density distributions without assumptions of self-similarity. 
Using their PDF fit formula for $M_{vir}$ at $z = 0.4$, we obtain $q=0.57\pm 0.17$.
These results match ours closely, and more recent independent analyses appear to be converging.
Note that the other simulations were DM-only, but \citet{Suto:2016hoj} found that non-sphericity is unaffected by baryonic physics beyond half of the virial radius, so it is reasonable to compare to our measurements. 

Several observational studies of weak-lensing have attempted to measure cluster halo ellipticity. In the Sloan Digital Sky Survey (SDSS), \cite{Evans:2008mp} found a mean projected axis ratio $\langle q\rangle=0.48^{+0.14}_{-0.09}$ in the redshift range $0.1 <z< 0.3$. 
By directly fitting 2D shear-maps with eNFW models, \cite{2010MNRAS.405.2215O} measured an mean projected axis ratio $\langle q\rangle=0.54\pm0.04$ for a sample of 18 X-ray luminous clusters in the redshift range $0.15<z<0.3$. 
\cite{Shin:2017rch} measured $\langle q\rangle= 0.56\pm0.09$ for 10,428 SDSS clusters.
These results are consistent with our measurement.
Intriguingly, \cite{Umetsu:2018ypz} measured the median projected axis-ratio of 20 high-mass galaxy clusters in the HST-CLASH survey to be $\langle q\rangle=0.67\pm0.07$, within a scale of 2\,Mpc$h^{-1}$. However, their measurement from the CLASH high-magnification subsample was $\langle q\rangle=0.55\pm0.11$, consistent with our results. This suggests a lensing selection bias towards halos that are more elliptical (in the plane of the sky as well as along a line of sight). In contrast, X-ray selected clusters tend to be relaxed clusters with rounder dark matter halo shapes. For clusters selected by the red sequence technique, it is more likely that they are elongated along the line of sight, causing an over-density of red galaxies in the projected sky-plane. Since our simulated cluster sample is selected by their high mass, with each halo projected along a random line-of-sight, we can only give the mass-selected mean halo shape.
For direct comparison with observational data, future theoretical predictions will need to take the selection function of the observed sample into effect.

Other shape measurement techniques are possible.
Studies using quadrupole estimators to quantify halo shape include \cite{Adhikari:2014hja,Clampitt:2015wea,vanUitert:2016guv,Shin:2017rch}. 
In particular, \cite{Clampitt:2015wea} developed a new estimator to measure the quadrupole weak-lensing signal from 70,000 SDSS Luminous Red Galaxies halos, and found a best-fit axis-ratio $\langle q\rangle\sim 0.78$. Their analysis assumes that dark matter perfectly aligns with light, so one potential systematic in their study is the possibility of light and dark matter misalignment. The determination of the orientation of each lens-source pair could become inaccurate due to this misalignment, and result in the dilution of the final stacked signal of the halo ellipticity.
Indeed, applying the misalignment distribution of \cite{Okumura:2008bm} to their measurement, they obtain $q\sim 0.6$, consistent with our results.

\subsection{Searches for filaments}
\label{sec:results:filaments}

\begin{figure*}
\centering
\includegraphics[width=0.68\textwidth]{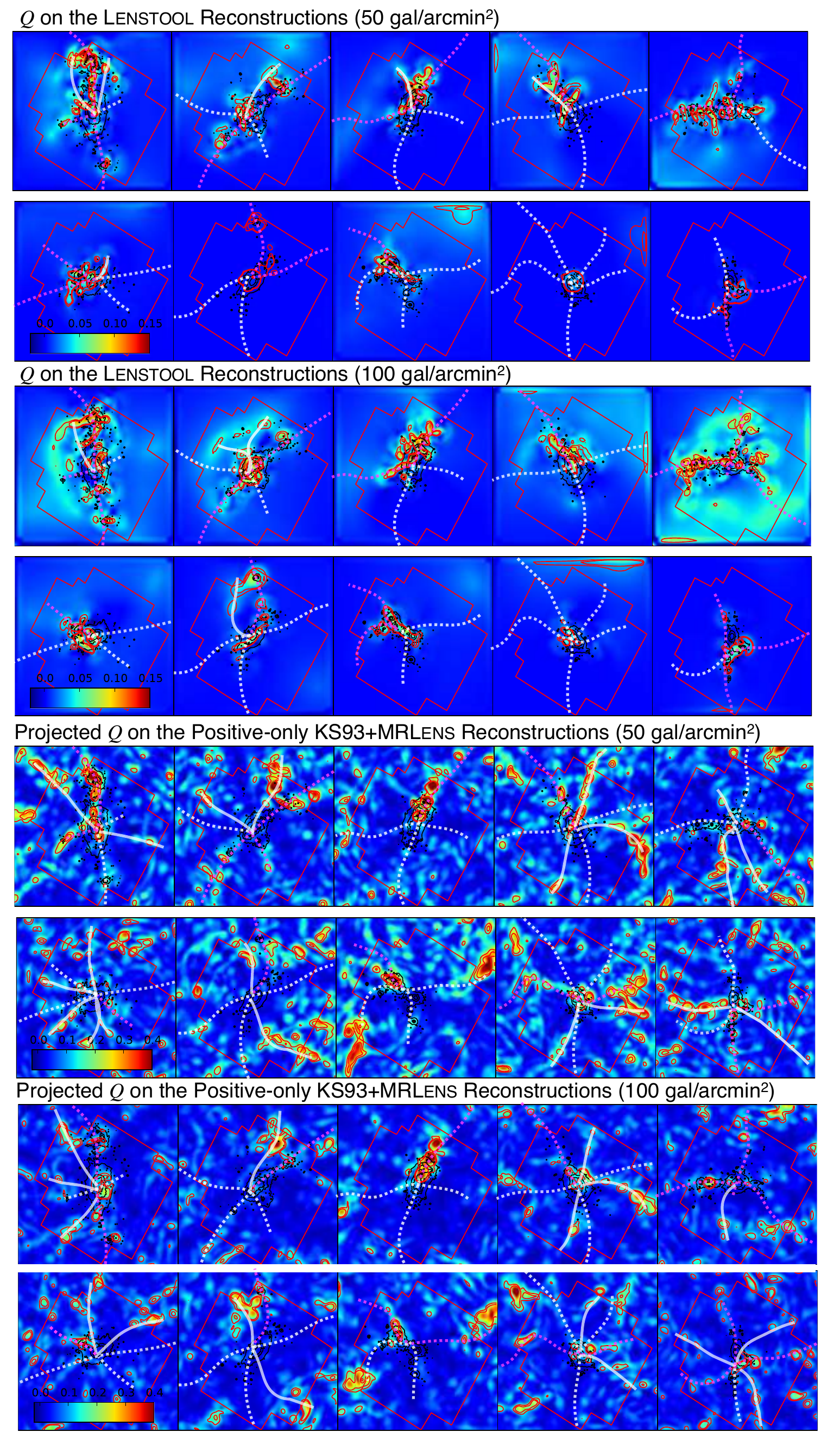}
\caption{Results for the filament search around 10 simulated clusters.
Colours show a linear combination of aperture multipole moments $Q$, calculated from the mass maps after subtracting their best-fit smooth component. Dotted lines show true filaments, reproduced from figure~\ref{fig:truekappa}; those identified successfully (with $Q_\mathrm{threshold}=3\sigma_Q$, see section~\ref{sec:method:filamentid}) are highlighted in magenta. Solid lines show false positive detections.
The top and second panel use mass maps created by \textsc{Lenstool} (including shape noise and LSS), with $50$\,arcmin$^{-2}$ and $100$\,arcmin$^{-2}$ source galaxies, respectively. The third and bottom panel show the phase-projected version of the filter applied to the positive-only KS93+\textsc{MRLens} mass map (with a different colour scale to the top two panels). In all panels, red contours show $Q=3\sigma_{Q}$ and $4\sigma_{Q}$, and black contours show the true mass distribution.}
\label{fig:Qmeasurements}
\end{figure*}

In the presence of galaxy shape noise and LSS noise, maps of our combination of aperture multipole moments $Q$ have lower signal-to-noise than maps of convergence $\kappa$ (figure~\ref{fig:Qmeasurements}; given the noise level, we show them only in linear scale, not logarithmic).
We quantify the noise level by defining $\sigma_Q$ as the standard deviation of all pixels in the final $Q$ map. 
Despite our attempt to eliminate isolated substructures from the $Q$ maps by combining different multipole moments, clusters 1, 2 and 5 contain sufficiently massive substructures to induce higher $Q$ than lower-density filaments.
Following the methodology in section~\ref{sec:method:filamentid}, we then search for filaments as extended regions with $Q>3\sigma_Q$ (illustrated in figure~\ref{fig:Qmeasurements}) or $Q>4\sigma_Q$.
Results for both are listed in table~\ref{tab:purity}.

In the default {\sc Lenstool} mass reconstructions, we find $\langle\sigma_Q\rangle=0.011$ and, with $Q_\mathrm{threshold}=3\sigma_Q$ we identify 17 of the 40 filaments (42.5$\%$ completeness), plus 5 false positive detections ($77.3\%$ purity).
Increasing the detection threshold to $4\sigma_Q$ removes all but one false detection, but finds only 12 real filaments.

Identifying filaments in the noisier KS93+\textsc{MRLens} mass reconstructions is much more difficult. To obtain useful results, we need to apply all three denoising strategies presented in Sect.~\ref{sec:method:denoising}.
We enlarge the apertures to $R_{\mathrm{max},0}$$=$$2\arcmin$,\   $R_{\mathrm{max},1}$$=$$R_{\mathrm{max},2}$$=$$2.5\arcmin$;
we replace negative convergence by zeros;
and we project all quadrupole and dipole moments in the radial direction.
In combination, these strategies reduce $\langle\sigma_Q\rangle$ from 0.11 to 0.06.
Filament identification statistics after this noise suppression are listed in table~\ref{tab:purity}. 
At $3\sigma_Q$ detection threshold, we identify 15 of the 40 filaments ($37.5\%$ completeness), but also 21 false positive detections ($41.7\%$ purity).

\begin{table}
\centering
\begin{tabular}{lrcccc}
\midrule
\midrule
\multicolumn{2}{r}{\multirow{2}{32mm}{~~~~~~~~~~~~Galaxy number\\ ~~~~~~~~~density [arcmin$^{-2}$]}} & \multicolumn{2}{|c|}{\textbf{Purity}}                                        & \multicolumn{2}{c}{\textbf{Completeness}} \\
&  & \multicolumn{1}{|c}{$3\,\sigma_Q$} & \multicolumn{1}{c|}{$4\,\sigma_Q$} & $3\,\sigma_Q$     & $4\,\sigma_Q$     \\
\midrule
 &20&  $35.0\%$    &  $40.0\%$  &  $50.0\%$  & $35.0\%$   \\
\textsc{\textbf{KS93+MRLens}} & 50  & $41.7\%$  &  $44.4\%$  & $37.5\%$ &  $30.0\%$ \\
& 100 & $50.0\%$ & $57.9\%$  &    $42.5\%$& $27.5\%$              \\
\midrule
 & 20 & $76.0\%$  & $78.0\%$ & $40.0\%$& $27.5\%$ \\
\textsc{\textbf{Lenstool}} & 50 & $77.3\%$  & $92.3\%$  &  $42.5\%$  & $30.0\%$  \\
 & 100 & $81.8\%$  & $93.3\%$ & $45.0\%$ & $35.0\%$ \\            
\midrule
\midrule
\end{tabular}
\caption{Filament identification efficiency at $3\sigma$ or $4\sigma$ detection significance, from multipole aperture moments in mass maps created by KS93+\textsc{MRLens} or \textsc{Lenstool}, assuming different densities of weakly lensed galaxies. 
Completeness indicates the fraction of the 40 real filaments (see section~\ref{sec:data:filaments}) that are successfully identified.
Purity indicates the fraction of the identified filaments that are real.
\label{tab:purity}}
\end{table}

Most of the false-positive filament detections are caused by galaxy shape noise.
Repeating the KS93+\textsc{MRLens} analysis with only shape noise yields a $Q$ map with $\sigma_Q=0.058$; with only LSS noise, it is $\sigma_Q=0.033$.
Because shape noise is apparently so dominant, we also investigate the effect of different survey strategies on the success of filament identification.
We simulate ground-based observations, which typically resolve the shapes of only 20\,galaxies armin$^{-2}$, and extremely deep space-based observations that resolve $\sim100$\,galaxies armin$^{-2}$ (we assume all faint galaxies have constant intrinsic shape noise, as suggested by figure~17 of \citealt{2007ApJS..172..219L}).
With these catalogues, we repeat the whole analysis: including the mass reconstruction and filament search (table~\ref{tab:purity}).
The low purity and high completeness of KS93+\textsc{MRLens} with $20$\,arcmin$^{-2}$ source galaxy is because the $Q$ maps are filled with random noise peaks that mimic the filament signals. Some radial directions defined by the alignment of noise peaks match the true filament direction by chance and thus boost the completeness in spite of low purity. Since these maps are not informative, we show only those $Q$ measurements using $100$\,arcmin$^{-2}$ source galaxies in figure~\ref{fig:Qmeasurements}.
The performance of {\sc Lenstool} reconstructions with deep space-based data is impressive: thanks to the prior assumption of looking harder where there are galaxies, it finds 18 filaments around 10 clusters ($45\%$ completeness) with $82\%$ purity. 
Recall that, even with noise-free data (section~\ref{sec:method:filamentid}), the maximum completeness with the multipole moment technique was 55\%.
In general, we find that {\sc Lenstool} is most appropriate for filament searches. Applied to future deep space-based surveys, the multipole moment technique should detect one or two filaments around most clusters.

\section{Conclusions}
\label{sec:conclusion}

High-precision calibration of weak-lensing mass reconstruction techniques will be essential for the next generation of space-based surveys. Understanding methods' performance in different systems (such as non-linear structures or stacked clusters), and quantifying any biases they introduce, will help identify the optimal method for each scientific analysis.

In this paper, we simulate mock observations of ten galaxy clusters from the BAHAMAS cosmological simulation. We use their known distribution of mass $\num{4e14} < M_{200} / \msun < \num{2e15}$ to test two mass mapping methods: (1) direct KS93 inversion from lensing shear observations to the projected mass distribution, which is then denoised using \textsc{MRLens}; (2) the forward-fitting \textsc{Lenstool} technique that uses a Bayesian MCMC sampler to fit the distribution of mass in a multi-scale grid. 
Any mass reconstruction method must interpolate the finite resolution in a shear catalogue that samples the shear field  
only along the lines of sight to galaxies. 

We find that \textsc{MRLens} is particularly efficient at suppressing noise owing to the diverse intrinsic shapes of background galaxies, whilst retaining signal from statistically significant structures on all scales. In a typical cluster field, it reduces total noise $\sigma_{\kappa}$ from $0.088\pm0.001$ to $0.026\pm0.001$. The \textsc{KS93+MRLens} method will be appropriate for use on stacked observations of a large number of galaxy clusters. However, it has no knowledge of cluster physics, and its noise suppression via smoothing softens the inferred central density profile. 
At large projected radii, $R>1$\,Mpc, noise in the map of an individual cluster becomes dominated by unrelated structures at different redshifts, projected along adjacent lines of sight.

\textsc{Lenstool} incorporates physical knowledge of galaxy clusters by imposing strong priors on the distribution of mass. For example, it preserves central cusps. The method is more aggressive in denoising the reconstructed convergence field, achieving $\sigma_{\kappa}=0.015\pm0.004$.
By adjusting the grid's adaptive resolution, it is also possible to suppress the spurious signal from unrelated, isolated structures at different redshifts, once they have been identified via multiband photometry or spectroscopy. 
We find that this method is well-suited to reconstructions of individual clusters, or measurements of low signal-to-noise quantities, such as filaments. 

In its standard configuration however, we find that \textsc{Lenstool} biases a mass reconstruction at large distances from the centre of a cluster, by imposing a prior that the projected density everywhere in a field of view must be positive (relative to the mean density in the Universe). This bias will need to be managed carefully when statistical errors are reduced by averaging over a population of clusters: perhaps by reconfiguring the Bayesian optimisation engine. 
The standard configuration of \textsc{Lenstool} also forces the mass distribution in every grid point to be spherically symmetric. In a purely weak-lensing analysis, this leads to spuriously spherical cluster cores, even when the global mass distribution is well modelled. This issue is automatically solved and irrelevant if strong gravitational lensing information is available, and used to pre-fit the axis ratio of the core. In this weak lensing-only study, we adopt a simple solution by masking the central $R<35\arcsec$ regions of a weak-lensing-only reconstruction. This avoids modelling the central spherical core for halo shape measurement.

Based on the performance of these two methods, for an individual cluster, or measurements of highly nonlinear quantities such as filament detection, \textsc{Lenstool} is well-suited to applications that require as precise a reconstruction as possible. However, for high-precision analyses that stack many clusters, it would be necessary to drop \textsc{Lenstool}'s positive definite constraint to reduce bias of mass over-estimation.
By contrast, \textsc{KS93+MRlens} retains a higher level of noise, but the positive and negative fluctuations are preserved in a manner which can reduce bias in stacked measurements. 

We also develop a filter to search for filaments and measure their orientation.
The low density of filaments leads to low signal-to-noise in reconstructed maps, and they can rarely be stacked usefully.
To retain their individual signal whilst suppressing noise, 
we construct a linear combination of multipole moments. We explore two further strategies: (1) filtering on the orientations (complex phases) of higher-order moments, exploiting the prior knowledge that filaments typically extend radially out of from cluster halos, and (2) replacing with the mean density of the Universe those regions inferred to have (negative) less density, which are more likely to be noise than regions inferred to have (positive) higher density.
We find that it will be impossible to detect individual filaments using data from ground-based telescopes, and remains challenging with current space-based ({\em HST}) data.
However, we find that the dominant source of noise relevant to filament detection comes from lensed galaxies' intrinsic shapes. Deeper observations with the next generation of space-based telescopes will resolve more background galaxies, and efficiently beat down this noise. 
Our filtering method successfully finds 45\% of filaments with projected density $\Sigma>1.7\times10^{7}\,\rm{\msun/kpc^2}$ (with a false detection rate $<$20\%), when applied to mock observations at the depth of possible future surveys.

\section*{Acknowledgements}
We would like to thank anonymous referee for giving useful
comments and improving our manuscript. 
We are grateful to Ian McCarthy for sharing his BAHAMAS simulation data, and supporting its interpretation.
SIT is supported by Van Mildert College Trust PhD Scholarship. RM is supported by a Royal Society University Research Fellowship. MJ is supported by the United Kingdom Research and Innovation (UKRI) Future Leaders Fellowship `Using Cosmic Beasts to uncover the Nature of Dark Matter' (grant number MR/S017216/1) and the UK Science and Technology Facilities Council (grant number ST/P000541/1). 
AR is supported by the European Research Council (project ERCStG-716532-PUNCA).

This work was also supported by the UK Science and Technology Facilities Council (grant number ST/L00075X/1). It used the DiRAC Durham facility managed by the Institute for Computational Cosmology on behalf of the STFC DiRAC HPC Facility (www.dirac.ac.uk). The equipment was funded by BEIS capital funding via STFC capital grants ST/K00042X/1, ST/P002293/1 and ST/R002371/1, Durham University and STFC operations grant ST/R000832/1. DiRAC is part of the UK National e-Infrastructure. 

\section*{Data Availability}
The simulation data underlying this article are available from \url{https://www.astro.ljmu.ac.uk/~igm/BAHAMAS/}.

\bibliographystyle{mnras}
\bibliography{paper3}

\begin{thebibliography}{}
\makeatletter
\relax
\def\mn@urlcharsother{\let\do\@makeother \do\$\do\&\do\#\do\^\do\_\do\%\do\~}
\def\mn@doi{\begingroup\mn@urlcharsother \@ifnextchar [ {\mn@doi@}
  {\mn@doi@[]}}
\def\mn@doi@[#1]#2{\def\@tempa{#1}\ifx\@tempa\@empty \href
  {http://dx.doi.org/#2} {doi:#2}\else \href {http://dx.doi.org/#2} {#1}\fi
  \endgroup}
\def\mn@eprint#1#2{\mn@eprint@#1:#2::\@nil}
\def\mn@eprint@arXiv#1{\href {http://arxiv.org/abs/#1} {{\tt arXiv:#1}}}
\def\mn@eprint@dblp#1{\href {http://dblp.uni-trier.de/rec/bibtex/#1.xml}
  {dblp:#1}}
\def\mn@eprint@#1:#2:#3:#4\@nil{\def\@tempa {#1}\def\@tempb {#2}\def\@tempc
  {#3}\ifx \@tempc \@empty \let \@tempc \@tempb \let \@tempb \@tempa \fi \ifx
  \@tempb \@empty \def\@tempb {arXiv}\fi \@ifundefined
  {mn@eprint@\@tempb}{\@tempb:\@tempc}{\expandafter \expandafter \csname
  mn@eprint@\@tempb\endcsname \expandafter{\@tempc}}}

\bibitem[\protect\citeauthoryear{Adhikari, Chue  \& Dalal}{Adhikari
  et~al.}{2015}]{Adhikari:2014hja}
Adhikari S.,  Chue C. Y.~R.,   Dalal N.,  2015, \mn@doi [JCAP]
  {10.1088/1475-7516/2015/01/009}, 1501, 009

\bibitem[\protect\citeauthoryear{{Alpaslan} et~al.,}{{Alpaslan}
  et~al.}{2015}]{2015MNRAS.451.3249A}
{Alpaslan} M.,  et~al., 2015, \mn@doi [\mnras] {10.1093/mnras/stv1176}, \href
  {https://ui.adsabs.harvard.edu/abs/2015MNRAS.451.3249A} {451, 3249}

\bibitem[\protect\citeauthoryear{{Alpaslan} et~al.,}{{Alpaslan}
  et~al.}{2016}]{2016MNRAS.457.2287A}
{Alpaslan} M.,  et~al., 2016, \mn@doi [\mnras] {10.1093/mnras/stw134}, \href
  {https://ui.adsabs.harvard.edu/abs/2016MNRAS.457.2287A} {457, 2287}

\bibitem[\protect\citeauthoryear{{Arnouts} et~al.,}{{Arnouts}
  et~al.}{2007}]{2007A&A...476..137A}
{Arnouts} S.,  et~al., 2007, \mn@doi [\aap] {10.1051/0004-6361:20077632}, \href
  {http://adsabs.harvard.edu/abs/2007A%26A...476..137A} {476, 137}

\bibitem[\protect\citeauthoryear{{Bahcall} \& {Bode}}{{Bahcall} \&
  {Bode}}{2003}]{2003ApJ...588L...1B}
{Bahcall} N.~A.,  {Bode} P.,  2003, \mn@doi [\apjl] {10.1086/375503}, \href
  {https://ui.adsabs.harvard.edu/abs/2003ApJ...588L...1B} {588, L1}

\bibitem[\protect\citeauthoryear{{Bahcall} \& {Cen}}{{Bahcall} \&
  {Cen}}{1993}]{1993ApJ...407L..49B}
{Bahcall} N.~A.,  {Cen} R.,  1993, \mn@doi [\apjl] {10.1086/186803}, \href
  {https://ui.adsabs.harvard.edu/abs/1993ApJ...407L..49B} {407, L49}

\bibitem[\protect\citeauthoryear{{Bartelmann}}{{Bartelmann}}{1996}]{1996A&A...313..697B}
{Bartelmann} M.,  1996, \aap, \href
  {https://ui.adsabs.harvard.edu/abs/1996A%26A...313..697B} {313, 697}

\bibitem[\protect\citeauthoryear{{Bartelmann} \& {Maturi}}{{Bartelmann} \&
  {Maturi}}{2017}]{srev}
{Bartelmann} M.,  {Maturi} M.,  2017, \mn@doi [Scholarpedia]
  {10.4249/scholarpedia.32440}, \href
  {https://ui.adsabs.harvard.edu/abs/2017SchpJ..1232440B} {12, 32440}

\bibitem[\protect\citeauthoryear{Bradac et~al.,}{Bradac
  et~al.}{2006}]{Bradac:2006er}
Bradac M.,  et~al., 2006, \mn@doi [Astrophys. J.] {10.1086/508601}, 652, 937

\bibitem[\protect\citeauthoryear{Byrd, Lu, Nocedal  \& Zhu}{Byrd
  et~al.}{1995}]{53712fe04a3448cfb8598b14afab59b3}
Byrd R.,  Lu P.,  Nocedal J.,   Zhu C.,  1995, \mn@doi [SIAM Journal of
  Scientific Computing] {10.1137/0916069}, 16, 1190

\bibitem[\protect\citeauthoryear{Chiu, Umetsu, Sereno, Ettori, Meneghetti,
  Merten, Sayers  \& Zitrin}{Chiu et~al.}{2018}]{Chiu:2018gok}
Chiu I.-N.,  Umetsu K.,  Sereno M.,  Ettori S.,  Meneghetti M.,  Merten J.,
  Sayers J.,   Zitrin A.,  2018, \mn@doi [Astrophys. J.]
  {10.3847/1538-4357/aac4a0}, 860, 126

\bibitem[\protect\citeauthoryear{Clampitt \& Jain}{Clampitt \&
  Jain}{2016}]{Clampitt:2015wea}
Clampitt J.,  Jain B.,  2016, \mn@doi [MNRAS] {10.1093/mnras/stw254}, 457, 4135

\bibitem[\protect\citeauthoryear{{Clowe} et~al.,}{{Clowe}
  et~al.}{2006}]{2006A&A...451..395C}
{Clowe} D.,  et~al., 2006, \mn@doi [\aap] {10.1051/0004-6361:20041787}, \href
  {https://ui.adsabs.harvard.edu/abs/2006A&A...451..395C} {451, 395}

\bibitem[\protect\citeauthoryear{{Crain} et~al.,}{{Crain}
  et~al.}{2009}]{2009MNRAS.399.1773C}
{Crain} R.~A.,  et~al., 2009, \mn@doi [\mnras]
  {10.1111/j.1365-2966.2009.15402.x}, \href
  {https://ui.adsabs.harvard.edu/abs/2009MNRAS.399.1773C} {399, 1773}

\bibitem[\protect\citeauthoryear{Despali, Giocoli, Bonamigo, Limousin  \&
  Tormen}{Despali et~al.}{2017}]{Despali:2016pkj}
Despali G.,  Giocoli C.,  Bonamigo M.,  Limousin M.,   Tormen G.,  2017,
  \mn@doi [MNRAS] {10.1093/mnras/stw3129}, 466, 181

\bibitem[\protect\citeauthoryear{Diemer \& Kravtsov}{Diemer \&
  Kravtsov}{2014}]{Diemer:2014xya}
Diemer B.,  Kravtsov A.~V.,  2014, \mn@doi [Astrophys. J.]
  {10.1088/0004-637X/789/1/1}, 789, 1

\bibitem[\protect\citeauthoryear{Dietrich, Schneider, Clowe, Romano-Diaz  \&
  Kerp}{Dietrich et~al.}{2005}]{Dietrich:2004kf}
Dietrich J.~P.,  Schneider P.,  Clowe D.,  Romano-Diaz E.,   Kerp J.,  2005,
  \mn@doi [\aap] {10.1051/0004-6361:20041523}, 440, 453

\bibitem[\protect\citeauthoryear{{Dietrich}, {Werner}, {Clowe}, {Finoguenov},
  {Kitching}, {Miller}  \& {Simionescu}}{{Dietrich}
  et~al.}{2012}]{2012Natur.487..202D}
{Dietrich} J.~P.,  {Werner} N.,  {Clowe} D.,  {Finoguenov} A.,  {Kitching} T.,
  {Miller} L.,   {Simionescu} A.,  2012, \mn@doi [\nat] {10.1038/nature11224},
  \href {http://adsabs.harvard.edu/abs/2012Natur.487..202D} {487, 202}

\bibitem[\protect\citeauthoryear{{Einasto} et~al.,}{{Einasto}
  et~al.}{2007}]{2007A&A...464..815E}
{Einasto} M.,  et~al., 2007, \mn@doi [\aap] {10.1051/0004-6361:20066456}, \href
  {https://ui.adsabs.harvard.edu/abs/2007A&A...464..815E} {464, 815}

\bibitem[\protect\citeauthoryear{{El{\'{\i}}asd{\'o}ttir}
  et~al.,}{{El{\'{\i}}asd{\'o}ttir} et~al.}{2007}]{eliasdottir2007}
{El{\'{\i}}asd{\'o}ttir} {\'A}.,  et~al., 2007, preprint, \href
  {http://adsabs.harvard.edu/abs/2007arXiv0710.5636E} {} (\mn@eprint {arXiv}
  {0710.5636})

\bibitem[\protect\citeauthoryear{Evans \& Bridle}{Evans \&
  Bridle}{2009}]{Evans:2008mp}
Evans A. K.~D.,  Bridle S.,  2009, \mn@doi [Astrophys. J.]
  {10.1088/0004-637X/695/2/1446}, 695, 1446

\bibitem[\protect\citeauthoryear{Fluri, Kacprzak, Lucchi, Refregier, Amara,
  Hofmann  \& Schneider}{Fluri et~al.}{2019}]{Fluri:2019qtp}
Fluri J.,  Kacprzak T.,  Lucchi A.,  Refregier A.,  Amara A.,  Hofmann T.,
  Schneider A.,  2019

\bibitem[\protect\citeauthoryear{{Gal{\'a}rraga-Espinosa}, {Aghanim}, {Langer},
  {Gouin}  \& {Malavasi}}{{Gal{\'a}rraga-Espinosa}
  et~al.}{2020}]{2020arXiv200309697G}
{Gal{\'a}rraga-Espinosa} D.,  {Aghanim} N.,  {Langer} M.,  {Gouin} C.,
  {Malavasi} N.,  2020, arXiv e-prints, \href
  {https://ui.adsabs.harvard.edu/abs/2020arXiv200309697G} {p. arXiv:2003.09697}

\bibitem[\protect\citeauthoryear{{Gavazzi}, {Mellier}, {Fort}, {Cuillandre}  \&
  {Dantel-Fort}}{{Gavazzi} et~al.}{2004}]{2004A&A...422..407G}
{Gavazzi} R.,  {Mellier} Y.,  {Fort} B.,  {Cuillandre} J.~C.,   {Dantel-Fort}
  M.,  2004, \mn@doi [\aap] {10.1051/0004-6361:20047109}, \href
  {https://ui.adsabs.harvard.edu/abs/2004A&A...422..407G} {422, 407}

\bibitem[\protect\citeauthoryear{{Golse} \& {Kneib}}{{Golse} \&
  {Kneib}}{2002}]{2002A&A...390..821G}
{Golse} G.,  {Kneib} J.~P.,  2002, \mn@doi [\aap] {10.1051/0004-6361:20020639},
  \href {https://ui.adsabs.harvard.edu/abs/2002A&A...390..821G} {390, 821}

\bibitem[\protect\citeauthoryear{{Gray}, {Taylor}, {Meisenheimer}, {Dye},
  {Wolf}  \& {Thommes}}{{Gray} et~al.}{2002}]{2002ApJ...568..141G}
{Gray} M.~E.,  {Taylor} A.~N.,  {Meisenheimer} K.,  {Dye} S.,  {Wolf} C.,
  {Thommes} E.,  2002, \mn@doi [\apj] {10.1086/338763}, \href
  {http://adsabs.harvard.edu/abs/2002ApJ...568..141G} {568, 141}

\bibitem[\protect\citeauthoryear{{Guzzo} et~al.,}{{Guzzo}
  et~al.}{2007}]{2007ApJS..172..254G}
{Guzzo} L.,  et~al., 2007, \mn@doi [\apjs] {10.1086/516588}, \href
  {https://ui.adsabs.harvard.edu/abs/2007ApJS..172..254G} {172, 254}

\bibitem[\protect\citeauthoryear{{Herbonnet} et~al.,}{{Herbonnet}
  et~al.}{2019}]{2019arXiv191204414H}
{Herbonnet} R.,  et~al., 2019, arXiv e-prints, \href
  {https://ui.adsabs.harvard.edu/abs/2019arXiv191204414H} {p. arXiv:1912.04414}

\bibitem[\protect\citeauthoryear{Heymans et~al.}{Heymans
  et~al.}{2005}]{Heymans:2004zp}
Heymans C.,  et~al., 2005, \mn@doi [MNRAS] {10.1111/j.1365-2966.2005.09152.x},
  361, 160

\bibitem[\protect\citeauthoryear{{Heymans} et~al.,}{{Heymans}
  et~al.}{2008}]{2008MNRAS.385.1431H}
{Heymans} C.,  et~al., 2008, \mn@doi [\mnras]
  {10.1111/j.1365-2966.2008.12919.x}, \href
  {http://ukads.nottingham.ac.uk/abs/2008MNRAS.385.1431H} {385, 1431}

\bibitem[\protect\citeauthoryear{{Hinshaw} et~al.,}{{Hinshaw}
  et~al.}{2013}]{2013ApJS..208...19H}
{Hinshaw} G.,  et~al., 2013, \mn@doi [\apjs] {10.1088/0067-0049/208/2/19},
  \href {http://adsabs.harvard.edu/abs/2013ApJS..208...19H} {208, 19}

\bibitem[\protect\citeauthoryear{Ho, Bahcall  \& Bode}{Ho
  et~al.}{2006}]{Ho:2005ea}
Ho S.,  Bahcall N.,   Bode P.,  2006, \mn@doi [Astrophys. J.] {10.1086/505255},
  647, 8

\bibitem[\protect\citeauthoryear{{Hoekstra}}{{Hoekstra}}{2013}]{hrev}
{Hoekstra} H.,  2013, arXiv e-prints, \href
  {https://ui.adsabs.harvard.edu/abs/2013arXiv1312.5981H} {p. arXiv:1312.5981}

\bibitem[\protect\citeauthoryear{Hopkins, Bahcall  \& Bode}{Hopkins
  et~al.}{2005}]{Hopkins_2005}
Hopkins P.~F.,  Bahcall N.~A.,   Bode P.,  2005, \mn@doi [The Astrophysical
  Journal] {10.1086/425993}, 618, 1

\bibitem[\protect\citeauthoryear{{Jauzac} et~al.,}{{Jauzac}
  et~al.}{2012}]{2012MNRAS.426.3369J}
{Jauzac} M.,  et~al., 2012, \mn@doi [\mnras]
  {10.1111/j.1365-2966.2012.21966.x}, \href
  {http://adsabs.harvard.edu/abs/2012MNRAS.426.3369J} {426, 3369}

\bibitem[\protect\citeauthoryear{{Jauzac} et~al.,}{{Jauzac}
  et~al.}{2015a}]{2015MNRAS.446.4132J}
{Jauzac} M.,  et~al., 2015a, \mn@doi [\mnras] {10.1093/mnras/stu2425}, \href
  {http://adsabs.harvard.edu/abs/2015MNRAS.446.4132J} {446, 4132}

\bibitem[\protect\citeauthoryear{Jauzac et~al.}{Jauzac
  et~al.}{2015b}]{Jauzac:2014xwa}
Jauzac M.,  et~al., 2015b, \mn@doi [MNRAS] {10.1093/mnras/stv1402}, 452, 1437

\bibitem[\protect\citeauthoryear{Jauzac et~al.}{Jauzac
  et~al.}{2016}]{Jauzac:2016tjc}
Jauzac M.,  et~al., 2016, \mn@doi [MNRAS] {10.1093/mnras/stw2251}, 463, 3876

\bibitem[\protect\citeauthoryear{{Jee}, {Hughes}, {Menanteau}, {Sif{\'o}n},
  {Mandelbaum}, {Barrientos}, {Infante}  \& {Ng}}{{Jee}
  et~al.}{2014}]{2014ApJ...785...20J}
{Jee} M.~J.,  {Hughes} J.~P.,  {Menanteau} F.,  {Sif{\'o}n} C.,  {Mandelbaum}
  R.,  {Barrientos} L.~F.,  {Infante} L.,   {Ng} K.~Y.,  2014, \mn@doi [\apj]
  {10.1088/0004-637X/785/1/20}, \href
  {https://ui.adsabs.harvard.edu/abs/2014ApJ...785...20J} {785, 20}

\bibitem[\protect\citeauthoryear{Jing \& Suto}{Jing \&
  Suto}{2002}]{Jing:2002np}
Jing Y.~P.,  Suto Y.,  2002, \mn@doi [Astrophys. J.] {10.1086/341065}, 574, 538

\bibitem[\protect\citeauthoryear{{Jullo} \& {Kneib}}{{Jullo} \&
  {Kneib}}{2009}]{2009MNRAS.395.1319J}
{Jullo} E.,  {Kneib} J.-P.,  2009, \mn@doi [\mnras]
  {10.1111/j.1365-2966.2009.14654.x}, \href
  {http://adsabs.harvard.edu/abs/2009MNRAS.395.1319J} {395, 1319}

\bibitem[\protect\citeauthoryear{{Jullo}, {Kneib}, {Limousin},
  {El{\'{\i}}asd{\'o}ttir}, {Marshall}  \& {Verdugo}}{{Jullo}
  et~al.}{2007}]{2007NJPh....9..447J}
{Jullo} E.,  {Kneib} J.-P.,  {Limousin} M.,  {El{\'{\i}}asd{\'o}ttir} {\'A}.,
  {Marshall} P.~J.,   {Verdugo} T.,  2007, \mn@doi [New Journal of Physics]
  {10.1088/1367-2630/9/12/447}, \href
  {http://adsabs.harvard.edu/abs/2007NJPh....9..447J} {9, 447}

\bibitem[\protect\citeauthoryear{{Jullo}, {Pires}, {Jauzac}  \&
  {Kneib}}{{Jullo} et~al.}{2014}]{2014MNRAS.437.3969J}
{Jullo} E.,  {Pires} S.,  {Jauzac} M.,   {Kneib} J.-P.,  2014, \mn@doi [\mnras]
  {10.1093/mnras/stt2207}, \href
  {http://adsabs.harvard.edu/abs/2014MNRAS.437.3969J} {437, 3969}

\bibitem[\protect\citeauthoryear{{Kaiser} \& {Squires}}{{Kaiser} \&
  {Squires}}{1993}]{1993ApJ...404..441K}
{Kaiser} N.,  {Squires} G.,  1993, \mn@doi [\apj] {10.1086/172297}, \href
  {http://adsabs.harvard.edu/abs/1993ApJ...404..441K} {404, 441}

\bibitem[\protect\citeauthoryear{Kaiser, Wilson, Luppino, Kofman, Gioia,
  Metzger  \& Dahle}{Kaiser et~al.}{1998}]{Kaiser:1998ja}
Kaiser N.,  Wilson G.,  Luppino G.,  Kofman L.,  Gioia I.,  Metzger M.,   Dahle
  H.,  1998

\bibitem[\protect\citeauthoryear{{Kassiola} \& {Kovner}}{{Kassiola} \&
  {Kovner}}{1993}]{1993ApJ...417..450K}
{Kassiola} A.,  {Kovner} I.,  1993, \mn@doi [\apj] {10.1086/173325}, \href
  {https://ui.adsabs.harvard.edu/abs/1993ApJ...417..450K} {417, 450}

\bibitem[\protect\citeauthoryear{{Kilbinger}}{{Kilbinger}}{2015}]{kirev}
{Kilbinger} M.,  2015, \mn@doi [Reports on Progress in Physics]
  {10.1088/0034-4885/78/8/086901}, \href
  {https://ui.adsabs.harvard.edu/abs/2015RPPh...78h6901K} {78, 086901}

\bibitem[\protect\citeauthoryear{{Kneib} \& {Natarajan}}{{Kneib} \&
  {Natarajan}}{2011}]{knrev}
{Kneib} J.-P.,  {Natarajan} P.,  2011, \mn@doi [\aapr]
  {10.1007/s00159-011-0047-3}, \href
  {https://ui.adsabs.harvard.edu/abs/2011A&ARv..19...47K} {19, 47}

\bibitem[\protect\citeauthoryear{{Kneib}, {Ellis}, {Smail}, {Couch}  \&
  {Sharples}}{{Kneib} et~al.}{1996}]{1996ApJ...471..643K}
{Kneib} J.-P.,  {Ellis} R.~S.,  {Smail} I.,  {Couch} W.~J.,   {Sharples} R.~M.,
   1996, \mn@doi [\apj] {10.1086/177995}, \href
  {http://adsabs.harvard.edu/abs/1996ApJ...471..643K} {471, 643}

\bibitem[\protect\citeauthoryear{Krolewski et~al.}{Krolewski
  et~al.}{2018}]{Krolewski:2017jsm}
Krolewski A.,  et~al., 2018, \mn@doi [Astrophys. J.]
  {10.3847/1538-4357/aac829}, 861, 60

\bibitem[\protect\citeauthoryear{{Kuutma}, {Tamm}  \& {Tempel}}{{Kuutma}
  et~al.}{2017}]{2017A&A...600L...6K}
{Kuutma} T.,  {Tamm} A.,   {Tempel} E.,  2017, \mn@doi [\aap]
  {10.1051/0004-6361/201730526}, \href
  {https://ui.adsabs.harvard.edu/abs/2017A&A...600L...6K} {600, L6}

\bibitem[\protect\citeauthoryear{{Laureijs} et~al.,}{{Laureijs}
  et~al.}{2011}]{2011arXiv1110.3193L}
{Laureijs} R.,  et~al., 2011, arXiv e-prints, \href
  {https://ui.adsabs.harvard.edu/abs/2011arXiv1110.3193L} {p. arXiv:1110.3193}

\bibitem[\protect\citeauthoryear{{Leauthaud} et~al.,}{{Leauthaud}
  et~al.}{2007}]{2007ApJS..172..219L}
{Leauthaud} A.,  et~al., 2007, \mn@doi [\apjs] {10.1086/516598}, \href
  {http://adsabs.harvard.edu/abs/2007ApJS..172..219L} {172, 219}

\bibitem[\protect\citeauthoryear{{Leonard}, {Pires}  \& {Starck}}{{Leonard}
  et~al.}{2012}]{2012MNRAS.423.3405L}
{Leonard} A.,  {Pires} S.,   {Starck} J.-L.,  2012, \mn@doi [\mnras]
  {10.1111/j.1365-2966.2012.21133.x}, \href
  {https://ui.adsabs.harvard.edu/abs/2012MNRAS.423.3405L} {423, 3405}

\bibitem[\protect\citeauthoryear{{Leonard}, {Lanusse}  \& {Starck}}{{Leonard}
  et~al.}{2015}]{2015MNRAS.449.1146L}
{Leonard} A.,  {Lanusse} F.,   {Starck} J.-L.,  2015, \mn@doi [\mnras]
  {10.1093/mnras/stv386}, \href
  {https://ui.adsabs.harvard.edu/abs/2015MNRAS.449.1146L} {449, 1146}

\bibitem[\protect\citeauthoryear{{Limousin}, {Kneib}  \&
  {Natarajan}}{{Limousin} et~al.}{2005}]{2005MNRAS.356..309L}
{Limousin} M.,  {Kneib} J.-P.,   {Natarajan} P.,  2005, \mn@doi [\mnras]
  {10.1111/j.1365-2966.2004.08449.x}, \href
  {http://adsabs.harvard.edu/abs/2005MNRAS.356..309L} {356, 309}

\bibitem[\protect\citeauthoryear{Liu, Hao, Wang  \& Yang}{Liu
  et~al.}{2019}]{Liu:2019mfb}
Liu C.,  Hao L.,  Wang H.,   Yang X.,  2019, \mn@doi [Astrophys. J.]
  {10.3847/1538-4357/ab1ea0}, 878, 69

\bibitem[\protect\citeauthoryear{{Mao}, {Wang}, {Frenk}, {Gao}, {Li}, {Wang},
  {Cao}  \& {Li}}{{Mao} et~al.}{2018}]{2018MNRAS.478L..34M}
{Mao} T.-X.,  {Wang} J.,  {Frenk} C.~S.,  {Gao} L.,  {Li} R.,  {Wang} Q.,
  {Cao} X.,   {Li} M.,  2018, \mn@doi [\mnras] {10.1093/mnrasl/sly069}, \href
  {https://ui.adsabs.harvard.edu/abs/2018MNRAS.478L..34M} {478, L34}

\bibitem[\protect\citeauthoryear{Martinet et~al.,}{Martinet
  et~al.}{2016}]{Martinet:2016ind}
Martinet N.,  et~al., 2016, \mn@doi [\aap] {10.1051/0004-6361/201526444}, 590,
  A69

\bibitem[\protect\citeauthoryear{Martizzi, Vogelsberger, Torrey, Pillepich,
  Hansen, Marinacci  \& Hernquist}{Martizzi et~al.}{2019}]{Martizzi:2019sax}
Martizzi D.,  Vogelsberger M.,  Torrey P.,  Pillepich A.,  Hansen S.~H.,
  Marinacci F.,   Hernquist L.,  2019

\bibitem[\protect\citeauthoryear{Massey et~al.}{Massey
  et~al.}{2007a}]{Massey:2007gh}
Massey R.,  et~al., 2007a, \mn@doi [Astrophys. J. Suppl.] {10.1086/516599},
  172, 239

\bibitem[\protect\citeauthoryear{{Massey} et~al.,}{{Massey}
  et~al.}{2007b}]{2007Natur.445..286M}
{Massey} R.,  et~al., 2007b, \mn@doi [\nat] {10.1038/nature05497}, \href
  {http://adsabs.harvard.edu/abs/2007Natur.445..286M} {445, 286}

\bibitem[\protect\citeauthoryear{{Massey}, {Kitching}  \& {Richard}}{{Massey}
  et~al.}{2010}]{mrev}
{Massey} R.,  {Kitching} T.,   {Richard} J.,  2010, \mn@doi [Reports on
  Progress in Physics] {10.1088/0034-4885/73/8/086901}, \href
  {https://ui.adsabs.harvard.edu/abs/2010RPPh...73h6901M} {73, 086901}

\bibitem[\protect\citeauthoryear{{McCarthy}, {Schaye}, {Bird}  \& {Le
  Brun}}{{McCarthy} et~al.}{2017}]{2017MNRAS.465.2936M}
{McCarthy} I.~G.,  {Schaye} J.,  {Bird} S.,   {Le Brun} A.~M.~C.,  2017,
  \mn@doi [\mnras] {10.1093/mnras/stw2792}, \href
  {http://adsabs.harvard.edu/abs/2017MNRAS.465.2936M} {465, 2936}

\bibitem[\protect\citeauthoryear{{McCarthy}, {Bird}, {Schaye},
  {Harnois-Deraps}, {Font}  \& {van Waerbeke}}{{McCarthy}
  et~al.}{2018}]{2018MNRAS.476.2999M}
{McCarthy} I.~G.,  {Bird} S.,  {Schaye} J.,  {Harnois-Deraps} J.,  {Font}
  A.~S.,   {van Waerbeke} L.,  2018, \mn@doi [\mnras] {10.1093/mnras/sty377},
  \href {http://adsabs.harvard.edu/abs/2018MNRAS.476.2999M} {476, 2999}

\bibitem[\protect\citeauthoryear{McClintock et~al.}{McClintock
  et~al.}{2019}]{McClintock:2018bxh}
McClintock T.,  et~al., 2019, \mn@doi [Mon. Not. Roy. Astron. Soc.]
  {10.1093/mnras/sty2711}, 482, 1352

\bibitem[\protect\citeauthoryear{{Mead}, {King}  \& {McCarthy}}{{Mead}
  et~al.}{2010}]{2010MNRAS.401.2257M}
{Mead} J. M.~G.,  {King} L.~J.,   {McCarthy} I.~G.,  2010, \mn@doi [\mnras]
  {10.1111/j.1365-2966.2009.15840.x}, \href
  {https://ui.adsabs.harvard.edu/abs/2010MNRAS.401.2257M} {401, 2257}

\bibitem[\protect\citeauthoryear{{Medezinski} et~al.,}{{Medezinski}
  et~al.}{2018}]{2018PASJ...70S..28M}
{Medezinski} E.,  et~al., 2018, \mn@doi [\pasj] {10.1093/pasj/psx128}, \href
  {https://ui.adsabs.harvard.edu/abs/2018PASJ...70S..28M} {70, S28}

\bibitem[\protect\citeauthoryear{Merten, Cacciato, Meneghetti, Mignone  \&
  Bartelmann}{Merten et~al.}{2009}]{Merten:2008qf}
Merten J.,  Cacciato M.,  Meneghetti M.,  Mignone C.,   Bartelmann M.,  2009,
  \mn@doi [\aap] {10.1051/0004-6361/200810372}, 500, 681

\bibitem[\protect\citeauthoryear{{Merten} et~al.,}{{Merten}
  et~al.}{2011}]{2011MNRAS.417..333M}
{Merten} J.,  et~al., 2011, \mn@doi [\mnras]
  {10.1111/j.1365-2966.2011.19266.x}, \href
  {https://ui.adsabs.harvard.edu/abs/2011MNRAS.417..333M} {417, 333}

\bibitem[\protect\citeauthoryear{{Merten} et~al.,}{{Merten}
  et~al.}{2015}]{2015ApJ...806....4M}
{Merten} J.,  et~al., 2015, \mn@doi [\apj] {10.1088/0004-637X/806/1/4}, \href
  {https://ui.adsabs.harvard.edu/abs/2015ApJ...806....4M} {806, 4}

\bibitem[\protect\citeauthoryear{Miyatake et~al.}{Miyatake
  et~al.}{2019}]{Miyatake:2018lpb}
Miyatake H.,  et~al., 2019, \mn@doi [Astrophys. J.] {10.3847/1538-4357/ab0af0},
  875, 63

\bibitem[\protect\citeauthoryear{{Moran}, {Ellis}, {Treu}, {Smith}, {Rich}  \&
  {Smail}}{{Moran} et~al.}{2007}]{2007ApJ...671.1503M}
{Moran} S.~M.,  {Ellis} R.~S.,  {Treu} T.,  {Smith} G.~P.,  {Rich} R.~M.,
  {Smail} I.,  2007, \mn@doi [\apj] {10.1086/522303}, \href
  {https://ui.adsabs.harvard.edu/abs/2007ApJ...671.1503M} {671, 1503}

\bibitem[\protect\citeauthoryear{{More}, {Kravtsov}, {Dalal}  \&
  {Gottl{\"o}ber}}{{More} et~al.}{2011}]{2011ApJS..195....4M}
{More} S.,  {Kravtsov} A.~V.,  {Dalal} N.,   {Gottl{\"o}ber} S.,  2011, \mn@doi
  [\apjs] {10.1088/0067-0049/195/1/4}, \href
  {http://adsabs.harvard.edu/abs/2011ApJS..195....4M} {195, 4}

\bibitem[\protect\citeauthoryear{Natarajan et~al.}{Natarajan
  et~al.}{2017}]{Natarajan:2017sbo}
Natarajan P.,  et~al., 2017, \mn@doi [MNRAS] {10.1093/mnras/stw3385}, 468, 1962

\bibitem[\protect\citeauthoryear{Navarro, Frenk  \& White}{Navarro
  et~al.}{1996}]{Navarro:1995iw}
Navarro J.~F.,  Frenk C.~S.,   White S. D.~M.,  1996, \mn@doi [Astrophys. J.]
  {10.1086/177173}, 462, 563

\bibitem[\protect\citeauthoryear{Navarro, Frenk  \& White}{Navarro
  et~al.}{1997}]{Navarro:1996gj}
Navarro J.~F.,  Frenk C.~S.,   White S. D.~M.,  1997, \mn@doi [Astrophys. J.]
  {10.1086/304888}, 490, 493

\bibitem[\protect\citeauthoryear{{Newman}, {Treu}, {Ellis}, {Sand }, {Nipoti},
  {Richard}  \& {Jullo}}{{Newman} et~al.}{2013}]{2013ApJ...765...24N}
{Newman} A.~B.,  {Treu} T.,  {Ellis} R.~S.,  {Sand } D.~J.,  {Nipoti} C.,
  {Richard} J.,   {Jullo} E.,  2013, \mn@doi [\apj]
  {10.1088/0004-637X/765/1/24}, \href
  {https://ui.adsabs.harvard.edu/abs/2013ApJ...765...24N} {765, 24}

\bibitem[\protect\citeauthoryear{{Newman}, {Ellis}  \& {Treu}}{{Newman}
  et~al.}{2015}]{2015ApJ...814...26N}
{Newman} A.~B.,  {Ellis} R.~S.,   {Treu} T.,  2015, \mn@doi [\apj]
  {10.1088/0004-637X/814/1/26}, \href
  {https://ui.adsabs.harvard.edu/abs/2015ApJ...814...26N} {814, 26}

\bibitem[\protect\citeauthoryear{{Nuza}, {Kitaura}, {He{\ss}}, {Libeskind}  \&
  {M{\"u}ller}}{{Nuza} et~al.}{2014}]{2014MNRAS.445..988N}
{Nuza} S.~E.,  {Kitaura} F.-S.,  {He{\ss}} S.,  {Libeskind} N.~I.,
  {M{\"u}ller} V.,  2014, \mn@doi [\mnras] {10.1093/mnras/stu1746}, \href
  {https://ui.adsabs.harvard.edu/abs/2014MNRAS.445..988N} {445, 988}

\bibitem[\protect\citeauthoryear{{Oguri}, {Takada}, {Okabe}  \&
  {Smith}}{{Oguri} et~al.}{2010}]{2010MNRAS.405.2215O}
{Oguri} M.,  {Takada} M.,  {Okabe} N.,   {Smith} G.~P.,  2010, \mn@doi [\mnras]
  {10.1111/j.1365-2966.2010.16622.x}, \href
  {https://ui.adsabs.harvard.edu/\#abs/2010MNRAS.405.2215O} {405, 2215}

\bibitem[\protect\citeauthoryear{{Okabe} \& {Smith}}{{Okabe} \&
  {Smith}}{2016}]{2016MNRAS.461.3794O}
{Okabe} N.,  {Smith} G.~P.,  2016, \mn@doi [\mnras] {10.1093/mnras/stw1539},
  \href {http://adsabs.harvard.edu/abs/2016MNRAS.461.3794O} {461, 3794}

\bibitem[\protect\citeauthoryear{Okumura, Jing  \& Li}{Okumura
  et~al.}{2009}]{Okumura:2008bm}
Okumura T.,  Jing Y.~P.,   Li C.,  2009, \mn@doi [Astrophys. J.]
  {10.1088/0004-637X/694/1/214}, 694, 214

\bibitem[\protect\citeauthoryear{Pandey \& Bharadwaj}{Pandey \&
  Bharadwaj}{2006}]{Pandey:2006rj}
Pandey B.,  Bharadwaj S.,  2006, \mn@doi [MNRAS]
  {10.1111/j.1365-2966.2006.10894.x}, 372, 827

\bibitem[\protect\citeauthoryear{{Pires}, {Starck}, {Amara}, {Teyssier},
  {R{\'e}fr{\'e}gier}  \& {Fadili}}{{Pires} et~al.}{2009}]{2009MNRAS.395.1265P}
{Pires} S.,  {Starck} J.~L.,  {Amara} A.,  {Teyssier} R.,  {R{\'e}fr{\'e}gier}
  A.,   {Fadili} J.,  2009, \mn@doi [\mnras]
  {10.1111/j.1365-2966.2009.14625.x}, \href
  {https://ui.adsabs.harvard.edu/abs/2009MNRAS.395.1265P} {395, 1265}

\bibitem[\protect\citeauthoryear{{Pires}, {Starck}, {Amara},
  {R{\'e}fr{\'e}gier}  \& {Teyssier}}{{Pires}
  et~al.}{2010}]{2010AIPC.1241.1118P}
{Pires} S.,  {Starck} J.~L.,  {Amara} A.,  {R{\'e}fr{\'e}gier} A.,   {Teyssier}
  R.,  2010, in {Alimi} J.-M.,  {Fu{\"o}zfa} A.,  eds,  American Institute of
  Physics Conference Series Vol. 1241, American Institute of Physics Conference
  Series. pp 1118--1127 (\mn@eprint {arXiv} {0904.2995}),
  \mn@doi{10.1063/1.3462608}

\bibitem[\protect\citeauthoryear{{Postman} et~al.,}{{Postman}
  et~al.}{2012}]{2012ApJS..199...25P}
{Postman} M.,  et~al., 2012, \mn@doi [\apjs] {10.1088/0067-0049/199/2/25},
  \href {https://ui.adsabs.harvard.edu/abs/2012ApJS..199...25P} {199, 25}

\bibitem[\protect\citeauthoryear{{Raghunathan}, {Holder}, {Bartlett}, {Patil},
  {Reichardt}  \& {Whitehorn}}{{Raghunathan}
  et~al.}{2019}]{2019JCAP...11..037R}
{Raghunathan} S.,  {Holder} G.~P.,  {Bartlett} J.~G.,  {Patil} S.,  {Reichardt}
  C.~L.,   {Whitehorn} N.,  2019, \mn@doi [\jcap]
  {10.1088/1475-7516/2019/11/037}, \href
  {https://ui.adsabs.harvard.edu/abs/2019JCAP...11..037R} {2019, 037}

\bibitem[\protect\citeauthoryear{Rehmann et~al.}{Rehmann
  et~al.}{2019}]{Rehmann:2018nis}
Rehmann R.~L.,  et~al., 2019, \mn@doi [MNRAS] {10.1093/mnras/stz817}, 486, 77

\bibitem[\protect\citeauthoryear{{Richard}, {Kneib}, {Ebeling}, {Stark},
  {Egami}  \& {Fiedler}}{{Richard} et~al.}{2011}]{2011MNRAS.414L..31R}
{Richard} J.,  {Kneib} J.-P.,  {Ebeling} H.,  {Stark} D.~P.,  {Egami} E.,
  {Fiedler} A.~K.,  2011, \mn@doi [\mnras] {10.1111/j.1745-3933.2011.01050.x},
  \href {https://ui.adsabs.harvard.edu/abs/2011MNRAS.414L..31R} {414, L31}

\bibitem[\protect\citeauthoryear{{Robertson}, {Harvey}, {Massey}, {Eke},
  {McCarthy}, {Jauzac}, {Li}  \& {Schaye}}{{Robertson}
  et~al.}{2019}]{2019MNRAS.488.3646R}
{Robertson} A.,  {Harvey} D.,  {Massey} R.,  {Eke} V.,  {McCarthy} I.~G.,
  {Jauzac} M.,  {Li} B.,   {Schaye} J.,  2019, \mn@doi [\mnras]
  {10.1093/mnras/stz1815}, \href
  {https://ui.adsabs.harvard.edu/abs/2019MNRAS.488.3646R} {488, 3646}

\bibitem[\protect\citeauthoryear{{Romualdez} et~al.,}{{Romualdez}
  et~al.}{2016}]{2016arXiv160802502R}
{Romualdez} L.~J.,  et~al., 2016, arXiv e-prints, \href
  {https://ui.adsabs.harvard.edu/abs/2016arXiv160802502R} {p. arXiv:1608.02502}

\bibitem[\protect\citeauthoryear{{Romualdez} et~al.,}{{Romualdez}
  et~al.}{2018}]{2018SPIE10702E..0RR}
{Romualdez} L.~J.,  et~al., 2018, in \procspie. p. 107020R (\mn@eprint {arXiv}
  {1807.02887}), \mn@doi{10.1117/12.2307754}

\bibitem[\protect\citeauthoryear{{Rozo} et~al.,}{{Rozo}
  et~al.}{2010}]{2010ApJ...708..645R}
{Rozo} E.,  et~al., 2010, \mn@doi [\apj] {10.1088/0004-637X/708/1/645}, \href
  {https://ui.adsabs.harvard.edu/abs/2010ApJ...708..645R} {708, 645}

\bibitem[\protect\citeauthoryear{{Salpeter}}{{Salpeter}}{1955}]{1955ApJ...121..161S}
{Salpeter} E.~E.,  1955, \mn@doi [\apj] {10.1086/145971}, \href
  {https://ui.adsabs.harvard.edu/abs/1955ApJ...121..161S} {121, 161}

\bibitem[\protect\citeauthoryear{Schaye et~al.}{Schaye
  et~al.}{2015}]{Schaye:2014tpa}
Schaye J.,  et~al., 2015, \mn@doi [MNRAS] {10.1093/mnras/stu2058}, 446, 521

\bibitem[\protect\citeauthoryear{Schneider \& Bartelmann}{Schneider \&
  Bartelmann}{1997}]{Schneider:1996yy}
Schneider P.,  Bartelmann M.,  1997, \mn@doi [MNRAS] {10.1093/mnras/286.3.696},
  286, 696

\bibitem[\protect\citeauthoryear{Schrabback et~al.}{Schrabback
  et~al.}{2018}]{Schrabback:2016hac}
Schrabback T.,  et~al., 2018, \mn@doi [Mon. Not. Roy. Astron. Soc.]
  {10.1093/mnras/stx2666}, 474, 2635

\bibitem[\protect\citeauthoryear{{Schwinn}, {Jauzac}, {Baugh}, {Bartelmann},
  {Eckert}, {Harvey}, {Natarajan}  \& {Massey}}{{Schwinn}
  et~al.}{2017}]{Schwinn2017}
{Schwinn} J.,  {Jauzac} M.,  {Baugh} C.~M.,  {Bartelmann} M.,  {Eckert} D.,
  {Harvey} D.,  {Natarajan} P.,   {Massey} R.,  2017, \mn@doi [\mnras]
  {10.1093/mnras/stx277}, \href
  {https://ui.adsabs.harvard.edu/abs/2017MNRAS.467.2913S} {467, 2913}

\bibitem[\protect\citeauthoryear{{Sereno}, {Covone}, {Izzo}, {Ettori}, {Coupon}
   \& {Lieu}}{{Sereno} et~al.}{2017}]{2017MNRAS.472.1946S}
{Sereno} M.,  {Covone} G.,  {Izzo} L.,  {Ettori} S.,  {Coupon} J.,   {Lieu} M.,
   2017, \mn@doi [\mnras] {10.1093/mnras/stx2085}, \href
  {https://ui.adsabs.harvard.edu/abs/2017MNRAS.472.1946S} {472, 1946}

\bibitem[\protect\citeauthoryear{Shaw, Weller, Ostriker  \& Bode}{Shaw
  et~al.}{2006}]{Shaw:2005dy}
Shaw L.,  Weller J.,  Ostriker J.~P.,   Bode P.,  2006, \mn@doi [Astrophys. J.]
  {10.1086/505016}, 646, 815

\bibitem[\protect\citeauthoryear{Shin, Clampitt, Jain, Bernstein, Neil, Rozo
  \& Rykoff}{Shin et~al.}{2018}]{Shin:2017rch}
Shin T.-h.,  Clampitt J.,  Jain B.,  Bernstein G.,  Neil A.,  Rozo E.,   Rykoff
  E.,  2018, \mn@doi [MNRAS] {10.1093/mnras/stx3366}, 475, 2421

\bibitem[\protect\citeauthoryear{{Smith}, {Kneib}, {Smail}, {Mazzotta},
  {Ebeling}  \& {Czoske}}{{Smith} et~al.}{2005}]{2005MNRAS.359..417S}
{Smith} G.~P.,  {Kneib} J.-P.,  {Smail} I.,  {Mazzotta} P.,  {Ebeling} H.,
  {Czoske} O.,  2005, \mn@doi [\mnras] {10.1111/j.1365-2966.2005.08911.x},
  \href {https://ui.adsabs.harvard.edu/abs/2005MNRAS.359..417S} {359, 417}

\bibitem[\protect\citeauthoryear{{Smith} et~al.,}{{Smith}
  et~al.}{2010}]{2010MNRAS.409..169S}
{Smith} G.~P.,  et~al., 2010, \mn@doi [\mnras]
  {10.1111/j.1365-2966.2010.17311.x}, \href
  {https://ui.adsabs.harvard.edu/abs/2010MNRAS.409..169S} {409, 169}

\bibitem[\protect\citeauthoryear{{Spergel} et~al.,}{{Spergel}
  et~al.}{2013}]{2013arXiv1305.5422S}
{Spergel} D.,  et~al., 2013, arXiv e-prints, \href
  {https://ui.adsabs.harvard.edu/abs/2013arXiv1305.5422S} {p. arXiv:1305.5422}

\bibitem[\protect\citeauthoryear{{Springel}, {White}, {Tormen}  \&
  {Kauffmann}}{{Springel} et~al.}{2001}]{2001MNRAS.328..726S}
{Springel} V.,  {White} S.~D.~M.,  {Tormen} G.,   {Kauffmann} G.,  2001,
  \mn@doi [\mnras] {10.1046/j.1365-8711.2001.04912.x}, \href
  {http://adsabs.harvard.edu/abs/2001MNRAS.328..726S} {328, 726}

\bibitem[\protect\citeauthoryear{Springel et~al.,}{Springel
  et~al.}{2005}]{Springel2005SimulationsOT}
Springel V.,  et~al., 2005, Nature, 435, 629

\bibitem[\protect\citeauthoryear{{Starck}, {Pires}  \&
  {R{\'e}fr{\'e}gier}}{{Starck} et~al.}{2006}]{2006A&A...451.1139S}
{Starck} J.-L.,  {Pires} S.,   {R{\'e}fr{\'e}gier} A.,  2006, \mn@doi [\aap]
  {10.1051/0004-6361:20052997}, \href
  {http://adsabs.harvard.edu/abs/2006A%26A...451.1139S} {451, 1139}

\bibitem[\protect\citeauthoryear{{Steinhardt} et~al.,}{{Steinhardt}
  et~al.}{2020}]{2020ApJS..247...64S}
{Steinhardt} C.~L.,  et~al., 2020, \mn@doi [\apjs] {10.3847/1538-4365/ab75ed},
  \href {https://ui.adsabs.harvard.edu/abs/2020ApJS..247...64S} {247, 64}

\bibitem[\protect\citeauthoryear{Suto, Kitayama, Nishimichi, Sasaki  \&
  Suto}{Suto et~al.}{2016}]{Suto:2016zqb}
Suto D.,  Kitayama T.,  Nishimichi T.,  Sasaki S.,   Suto Y.,  2016, \mn@doi
  [Publ. Astron. Soc. Jap.] {10.1093/pasj/psw088}, 68, 97

\bibitem[\protect\citeauthoryear{Suto, Peirani, Dubois, Kitayama, Nishimichi,
  Sasaki  \& Suto}{Suto et~al.}{2017}]{Suto:2016hoj}
Suto D.,  Peirani S.,  Dubois Y.,  Kitayama T.,  Nishimichi T.,  Sasaki S.,
  Suto Y.,  2017, \mn@doi [Publ. Astron. Soc. Jap.] {10.1093/pasj/psw118}, 69,
  14

\bibitem[\protect\citeauthoryear{{Tam} et~al.,}{{Tam}
  et~al.}{2020}]{2020arXiv200610156T}
{Tam} S.-I.,  et~al., 2020, arXiv e-prints, \href
  {https://ui.adsabs.harvard.edu/abs/2020arXiv200610156T} {p. arXiv:2006.10156}

\bibitem[\protect\citeauthoryear{Treu \& Ellis}{Treu \& Ellis}{2015}]{trev}
Treu T.,  Ellis R.~S.,  2015, \mn@doi [Contemporary Physics]
  {10.1080/00107514.2015.1006001}, 56, 17

\bibitem[\protect\citeauthoryear{Umetsu et~al.}{Umetsu
  et~al.}{2014}]{Umetsu:2014vna}
Umetsu K.,  et~al., 2014, \mn@doi [Astrophys. J.]
  {10.1088/0004-637X/795/2/163}, 795, 163

\bibitem[\protect\citeauthoryear{Umetsu et~al.,}{Umetsu
  et~al.}{2015}]{Umetsu:2015hda}
Umetsu K.,  et~al., 2015, \mn@doi [Astrophys. J.]
  {10.1088/0004-637X/806/2/207}, 806, 207

\bibitem[\protect\citeauthoryear{Umetsu et~al.}{Umetsu
  et~al.}{2018}]{Umetsu:2018ypz}
Umetsu K.,  et~al., 2018, \mn@doi [Astrophys. J.] {10.3847/1538-4357/aac3d9},
  860, 104

\bibitem[\protect\citeauthoryear{{Umetsu} et~al.,}{{Umetsu}
  et~al.}{2020}]{2020ApJ...890..148U}
{Umetsu} K.,  et~al., 2020, \mn@doi [\apj] {10.3847/1538-4357/ab6bca}, \href
  {https://ui.adsabs.harvard.edu/abs/2020ApJ...890..148U} {890, 148}

\bibitem[\protect\citeauthoryear{{Van Waerbeke} et~al.,}{{Van Waerbeke}
  et~al.}{2013}]{2013MNRAS.433.3373V}
{Van Waerbeke} L.,  et~al., 2013, \mn@doi [\mnras] {10.1093/mnras/stt971},
  \href {http://adsabs.harvard.edu/abs/2013MNRAS.433.3373V} {433, 3373}

\bibitem[\protect\citeauthoryear{{Warren}, {Quinn}, {Salmon}  \&
  {Zurek}}{{Warren} et~al.}{1992}]{1992ApJ...399..405W}
{Warren} M.~S.,  {Quinn} P.~J.,  {Salmon} J.~K.,   {Zurek} W.~H.,  1992,
  \mn@doi [\apj] {10.1086/171937}, \href
  {https://ui.adsabs.harvard.edu/abs/1992ApJ...399..405W} {399, 405}

\bibitem[\protect\citeauthoryear{{Weinberg}, {Bullock}, {Governato}, {Kuzio de
  Naray}  \& {Peter}}{{Weinberg} et~al.}{2015}]{2015PNAS..11212249W}
{Weinberg} D.~H.,  {Bullock} J.~S.,  {Governato} F.,  {Kuzio de Naray} R.,
  {Peter} A.~H.~G.,  2015, \mn@doi [Proceedings of the National Academy of
  Science] {10.1073/pnas.1308716112}, \href
  {https://ui.adsabs.harvard.edu/abs/2015PNAS..11212249W} {112, 12249}

\bibitem[\protect\citeauthoryear{{White} \& {Rees}}{{White} \&
  {Rees}}{1978}]{1978MNRAS.183..341W}
{White} S.~D.~M.,  {Rees} M.~J.,  1978, \mn@doi [\mnras]
  {10.1093/mnras/183.3.341}, \href
  {http://adsabs.harvard.edu/abs/1978MNRAS.183..341W} {183, 341}

\bibitem[\protect\citeauthoryear{{White}, {Cohn}  \& {Smit}}{{White}
  et~al.}{2010}]{2010MNRAS.408.1818W}
{White} M.,  {Cohn} J.~D.,   {Smit} R.,  2010, \mn@doi [\mnras]
  {10.1111/j.1365-2966.2010.17248.x}, \href
  {https://ui.adsabs.harvard.edu/abs/2010MNRAS.408.1818W} {408, 1818}

\bibitem[\protect\citeauthoryear{{Wuyts}, {van Dokkum}, {Kelson}, {Franx}  \&
  {Illingworth}}{{Wuyts} et~al.}{2004}]{2004ApJ...605..677W}
{Wuyts} S.,  {van Dokkum} P.~G.,  {Kelson} D.~D.,  {Franx} M.,   {Illingworth}
  G.~D.,  2004, \mn@doi [\apj] {10.1086/381746}, \href
  {http://adsabs.harvard.edu/abs/2004ApJ...605..677W} {605, 677}

\bibitem[\protect\citeauthoryear{{Yuan} et~al.,}{{Yuan}
  et~al.}{2019}]{2019MNRAS.487.1315Y}
{Yuan} L.,  et~al., 2019, \mn@doi [\mnras] {10.1093/mnras/stz1266}, \href
  {https://ui.adsabs.harvard.edu/abs/2019MNRAS.487.1315Y} {487, 1315}

\bibitem[\protect\citeauthoryear{van Uitert et~al.}{van Uitert
  et~al.}{2017}]{vanUitert:2016guv}
van Uitert E.,  et~al., 2017, \mn@doi [MNRAS] {10.1093/mnras/stx344}, 467, 4131

\bibitem[\protect\citeauthoryear{{von der Linden} et~al.,}{{von der Linden}
  et~al.}{2014}]{2014MNRAS.439....2V}
{von der Linden} A.,  et~al., 2014, \mn@doi [\mnras] {10.1093/mnras/stt1945},
  \href {https://ui.adsabs.harvard.edu/abs/2014MNRAS.439....2V} {439, 2}

\makeatother
\end{thebibliography}

\bsp	
\label{lastpage}
\end{document}